# Donor and acceptor characteristics of native point defects in GaN

Zijuan Xie[1,2], Yu Sui[*,1], John Buckeridge[†,2], C. Richard A. Catlow[2], Thomas W. Keal[3], Paul Sherwood[3], Aron Walsh[4,5], Matthew R. Farrow[2], David O. Scanlon[2,6,7], Scott M. Woodley[2], and Alexey A. Sokol[‡,2]

[1] Department of Physics, Harbin Institute of Technology, Harbin, China
[2] Kathleen Lonsdale Materials Chemistry, Department of Chemistry, University College London, London, United Kingdom
[3] Daresbury Laboratory, Scientific Computing Department, STFC, Daresbury, Warrington, United Kingdom
[4] Department of Materials, Imperial College London, London, United Kingdom
[5] Department of Materials Science and Engineering, Yonsei University, Seoul, Korea
[6] Thomas Young Centre, University College London, London, United Kingdom
[7] Diamond Light Source Ltd., Diamond House, Harwell Science and Innovation Campus, Didcot, Oxfordshire, United Kingdom

E-mail: *suiyu@hit.edu.cn, †j.buckeridge@ucl.ac.uk, ‡a.sokol@ucl.ac.uk

**Abstract**

The semiconducting behaviour and optoelectronic response of gallium nitride is governed by point defect processes, which, despite many years of research, remain poorly understood. The key difficulty in the description of the dominant charged defects is determining a consistent position of the corresponding defect levels, which is difficult to derive using standard supercell calculations. In a complementary approach, we take advantage of the embedded cluster methodology that provides direct access to a common zero of the electrostatic potential for all point defects in all charge states. Charged defects polarise a host dielectric material with long-range forces that strongly affect the outcome of defect simulations; to account for the polarisation we couple embedding with the hybrid quantum mechanical/molecular mechanical (QM/MM) approach and investigate the structure, formation and ionisation energies, and equilibrium concentrations of native point defects in wurtzite GaN at a chemically accurate hybrid-density-functional-theory level. N vacancies are the most thermodynamically favourable native defects in GaN, which contribute to the n-type character of as-grown GaN but are not the main source, a result that is consistent with experiment. Our calculations show no native point defects can form thermodynamically stable acceptor states. GaN can be easily doped n-type, but, in equilibrium conditions at moderate temperatures acceptor dopants will be compensated by N vacancies and no significant hole concentrations will be observed, indicating non-equilibrium processes must dominate in p-type GaN. We identify spectroscopic signatures of native defects in the infrared, visible and ultraviolet luminescence ranges and complementary spectroscopies. Crucially, we calculate the effective-mass-like-state levels associated with electrons and holes bound in diffuse orbitals. These levels may be accessible in competition with more strongly-localised states in luminescence processes and allow the attribution of the observed 3.46 and 3.27 eV UV peaks in a broad range of GaN samples to the presence of N vacancies.

# 1. Introduction

GaN is a key material for many technologically important applications including blue-light emitting and laser diodes[1], high-power microelectronics[2], solar cells[3,4] and catalysis[5]. Although the properties of native defects in GaN govern many fundamental characteristics of the material, such properties are not well understood, despite much research, and, indeed, which defects dominate is still a major topic of debate. GaN belongs to the family of III–V compound semiconductors, which has been widely studied experimentally and theoretically[6]. It stabilises in the wurtzite phase under ambient conditions. The most important native point defects to consider can be proposed by analogy with other III-V systems[7,8], which are much better characterised, such as GaAs, where vacancies and antisites are thought to dominate[9,10]. A comprehensive study of native point defects in GaN should necessarily take into consideration lattice vacancies, self-interstitials, and antisites.

A conclusion drawn from many studies is that the *n*-type character of as-grown GaN is due to nitrogen vacancies[11,12], while others claim it is due to unwanted impurities[13,14]. The various photoluminescence peaks observed in undoped GaN have also been attributed to many different native defects[15]. Point defects in GaN can exist as either donors, which under certain conditions donate electrons to the conduction band, or acceptors, which donate holes to the valence band. In applications, the criterion for the donor or acceptor nature is that the activation energy should be comparable with the thermal energy. Of course, many defects can both donate and accept electrons, which results in the change of their charge states, and may have donor and acceptor levels far inside the band gap. Native defects can act as compensation, passivation, and recombination centres during doping, which makes understanding their properties a prerequisite to study impurities and extended defects. Due to the wide variety of defects, charge states, and ionisation levels possible, deciphering experimental results with respect to particular defect event is a difficult process requiring significant input from theory.

Two pioneering density functional theory (DFT) studies using the plane-wave supercell technique provided a comprehensive description of the structural and energetic properties of native point defects in wurtzite GaN, both performed over twenty years ago[16,17]. The compensation effect of native defects was stressed in the studies, and vacancies are identified as dominant in GaN, with N vacancies ($V_N$) dominating in *p*-type and Ga vacancies ($V_{Ga}$) in *n*-type GaN. Boguslawski *et al.* found that the $V_N$ introduces a shallow donor level and may be responsible for the n-type character of as-grown GaN[16], in agreement with earlier studies using tight-binding methods[18,19], while Neugebauer and Van de Walle argued that the defect has high formation energy under n-type conditions and thus could be excluded as the source of the n-type conductivity in as-grown GaN[17]. Both groups reported that the split-interstitial configuration was most favourable for the N interstitial ($N_i$). Neugebauer and Van de Walle further argued that the large mismatch in the covalent radii of Ga and N renders the antisites ($Ga_N$) and ($N_{Ga}$) and Ga interstitials ($Ga_i$) energetically less favourable than vacancies[17]. Boguslawski *et al.*, in contrast, found the concentration of $Ga_i$ ([$Ga_i$]) to be comparable to that of the vacancies (implying similar energies of formation)[16]. Finite-size effects resulting from employing the supercell method were not taken into consideration in either study due to restrictive computational loads and underdeveloped theory. Contemporary studies of native defects in the alternative, higher symmetry but metastable zinc blende (cubic) phase of GaN showed similar defect formation energies and electronic structures to those in wurtzite GaN[20–22].

The early results summarised above presented a baseline in our understanding of defect processes in GaN, which has been exploited in many experimental studies to interpret available or new data. With the development of new methods and increase in computational power, more sophisticated theoretical approaches have become more affordable in recent years and some of the early findings were challenged[23–34][35–44][45–48]. While those earlier studies did report defect levels, their comparison with experiment remained unclear. Indeed, the resulting carrier concentrations derived from the defect formation energies were considered the most appropriate properties to compare with experiment. In later studies, both adiabatic and optical levels were calculated and compared with the widely-available photoluminescence (PL) and other spectroscopic data. The introduction of hybrid density functionals to the plane-wave DFT approach, where a fraction of exact Hartree-Fock exchange is included in the density functional, allowed more accurate electronic structures to be calculated, although the exact implementation of this method varied from study to study.

Despite many theoretical and experimental investigations on native defects in GaN, there remain many controversies and unresolved problems, which is partly due to the difficulty of directly detecting dilute point defects in experiments. Examples include the sources of yellow, green, red, blue and ultraviolet PL peaks observed in many samples; the source of intrinsic *n*-type conductivity in GaN; the stability of the N antisite; and the concentration of Ga interstitials. The commonly used DFT supercell approach for charged defects calculations in semiconductors is well known to suffer from finite-size effects, i.e. spurious electron confinement (in common with cluster techniques) within the supercell and periodic image interactions, which are particularly problematic for charged defects. Moreover, the periodic boundary conditions do not allow for a unique definition of the energy reference level for different unit cells and charge states[49–52]. Many efforts have been made to resolve these issues[42,53–55]. The resulting variety in approaches is a plausible reason for the lack of consistency within different calculations and in comparison to experiments[42]. Another point is the choice of density functional, which has significant effect on the calculated energy levels[56]; the level of exact exchange employed in hybrid DFT has varied over many different studies, which also can account for the range of different results.

Here, we report a systematic hybrid quantum mechanical/molecular mechanical (QM/MM) embedded cluster study on the structural, electronic and optical properties of native defects in GaN. Our goal is to achieve a reliable, accurate and reproducible account of native point defects in wurtzite structured GaN using a method that avoids spurious interactions inherent in periodic models and thus provide a reference for comparison of computational and experimental studies of this material. By design, the embedded cluster method accounts for the short and long-range polarisation effects of charged defects on the host material; lacks periodic image interactions; and provides an unambiguous definition of the vacuum level, facilitating the calculation of ionisation energies with an absolute reference. Highly delocalised states, however, are poorly represented due to quantum confinement effects (within the QM region of the model), which, incidentally, can also enter the description of diffuse defect states by supercell methods. Efforts are underway to overcome this problem[57], but we note that, for sufficiently compact defect states, such as the majority of those studied here, confinement will present no significant problem, provided a suitable basis set is employed to describe the electronic levels. Moreover, the methodological framework provides access to a broader range of quantum chemical methods including hybrid density functionals than is commonly available in plane-wave DFT codes. We therefore employ two hybrid functionals, both of which have been designed to reproduce a comprehensive set of thermochemical data well, but one of which is also designed to reproduce kinetic data. We have indicated in a previous study that one of these functionals gives results in oxides very close to commonly applied hybrid functionals in plane-wave studies [68]; this functional then provides a benchmark for our calculations.

Our results show that vacancies will dominate in GaN, N antisites are highly unfavourable, $N_i$ stabilise in split interstitial configurations, $Ga_i$ preferentially occupy octahedral sites, and that thermodynamic concentrations of defects will be low apart from some special cases. We give a fully comprehensive list of all vertical transition energies, including both electron and hole ionisation, which may be used by experimentalists for comparison with various spectra. Our results broadly agree with earlier studies and can explain many experimental findings including photoluminescence, deep level transient spectroscopies and complementary techniques.

The remainder of the paper is structured as follows: in section 2 we provide details of our computational approach. In section 3 we present our results, beginning with the formation energies of each point defect studied, then the thermal transitions, defect concentrations, and the ionisation energies of all relevant charge states. Finally, we discuss implications of calculated defect energies for experimental observations and highlight remaining problems in defect assignment, and comment on the description of shallow states in hybrid QM/MM calculations.

## 2. Computational techniques

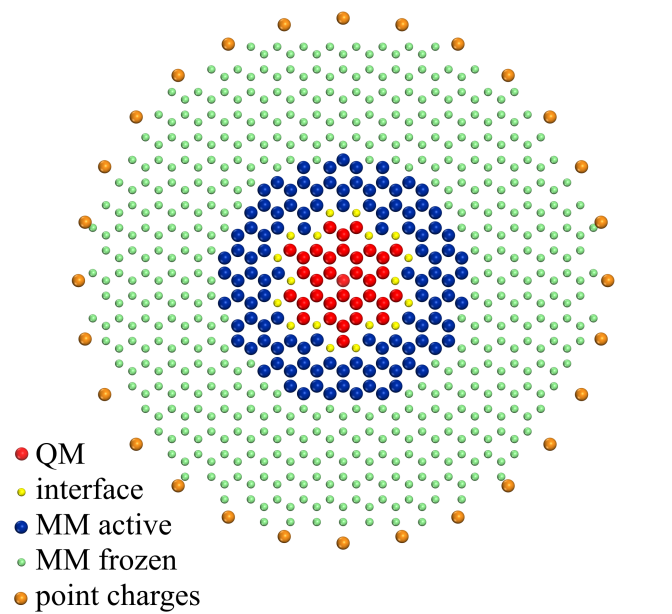

**Figure 1.** Schematic of the hybrid QM/MM embedded cluster model, represented by a two-dimensional cross section of the spherical cluster. The inner cluster (red spheres) is treated at a QM level of theory, and the surrounding atoms are modelled using an MM level of theory, divided into an active region (blue spheres) where ionic and electronic interactions are accounted for explicitly, and a frozen region (green spheres) where polarisation is treated a posteriori. Embedding pseudopotentials are placed on cation sites within the interface (yellow spheres) between the QM and MM regions, and the entire cluster is terminated by point charges (orange spheres) chosen to reproduce the Madelung potential of the infinite crystal.

The hybrid quantum mechanical/molecular mechanical (QM/MM) embedded cluster technique employed is designed to study localized states in ionic solids, where charged and strongly dipolar species predominantly interact via long-range electrostatic and short-range exchange forces[58,59]. This method and a plethora of closely related embedding approaches[60] split extended systems into the inner region, containing the central defect, described using molecular QM theories, and its surroundings, which are only slightly perturbed by the defect, modelled with MM approaches. Our choice of the QM methodology is DFT; the MM simulations employ the polarisable shell model interatomic potentials; and the interface between these two regions is based on cation centred semilocal pseudopotentials. The outer part of the MM region is held fixed at the pre-optimised geometry[61–63] and any polarisation here is treated *a posteriori* (see below). The entire cluster is surrounded by point charges, which are fitted in order to reproduce the Madelung potential of the infinite crystal within the inner QM region. A schematic of the model is shown in figure 1.

In our method, the inner cluster of 116 atoms of wurtzite GaN (pre-optimised using the MM model, see Refs. [61–63]) centred on the defect is treated using (i) the second-generation thermochemical hybrid exchange and correlation (XC) density functional B97-2[64], which is similar to those commonly used in recent plane-wave supercell calculations (21% exact exchange compared with 25% for PBE0[65] or HSE06[66]), (ii) the SBKJC small-core pseudopotentials on Ga[67] within the cluster and large-core refitted pseudopotentials[61,62] in the interface that provide a short-range contribution to the embedding potential on the defect, and (iii) the atomic basis set of def2-TZVP quality on N[68] and matching SBKJC basis on Ga[67]. For comparison, we use a second hybrid XC density functional employing 42% exact exchange (BB1k)[69], fitted to reproduce kinetic barriers and thermochemical data, which, given previous work on oxides[70–73], we expect gives a more accurate description of electron and hole localization than B97-2 – see also our earlier work on GaN in Refs. [62,74]. This QM region of radius 6.8 Å is embedded in an outer cluster of radius 30 Å, which is treated with an MM level of theory using two-body interatomic potentials parameterised to reproduce the wurtzite GaN bulk structure and physical properties[62,63]. A possible mis-match between the description of the MM region and the QM region could introduce strain erroneously in the model; however, both the MM model and QM hybrid density functionals we employ reproduce the low temperature experimental lattice parameters and bond lengths very well, resulting in a small mis-match (See the Supplemental Material for a demonstration of the reproduction of experimental bond lengths within the centre of the QM region using either functional). The method has been implemented in the ChemShell package[58,59,75] that employs Gamess-UK[76] for the QM and GULP[77] for MM single point energy and gradient calculations. Further technical details are discussed elsewhere[58,78]. This method has been previously applied to gain insights into the defect properties of a range of ionic systems[50,73,79–81] including earlier studies of GaN[57,62,82] and, moreover, the band alignment of polymorphs of TiO$_2$[62,83,84].

Defect levels of various defect charge states we report are calculated using their ionization energy, which consists of vertical (optical) or adiabatic (thermal, or thermodynamic) transitions of electrons to vacuum or, alternatively, the conduction band of a semiconductor. Periodic calculations confirm that the VBM in GaN is composed of N 2p states, which, although extended throughout the infinite periodic system, remain well localised on anion sites. Using the embedded cluster approach with a suitably terminated QM cluster (i.e. with cations as terminating ions and an appropriate embedding potential), we can model such states well. Therefore, by calculating the ionization energy of an electron from the bulk system, we can probe the VBM and determine its position relative to vacuum. Using the B97-2 functional (BB1k functional), we calculate the VBM to be 6.625 eV (7.340 eV) below the vacuum level, which is comparable with 6 eV as calculated using periodic slabs with the PBE-GGA[85] functional, 7.2 eV from surface *GW* calculations[86], and 6.8 eV from experiment[87]. The conduction band states, however, which are highly delocalised, are underbound in our embedded cluster calculations. To correct for this quantum confinement effect, we shift their position so that their minimum lies at an energy equal to the band gap above the VBM, relative to vacuum. For this procedure we use the experimental band gap value of 3.5 eV[88].

The formation energy of a defect $X(E_f[X])$ is determined from the relevant defect reaction as:
$$E_f[X] = \Delta E(X) - \sum_i n_i \mu_i + qE_f,$$
where $\Delta E(X)$ is the difference in energy between the embedded cluster with and without $X$, $n_i$ is the number of atoms of species $i$ added ($n_i > 0$) or subtracted ($n_i < 0$) in forming $X$, $\mu_i$ is the chemical potential of species $i$, $q$ is the charge of $X$, and $E_f$ is the Fermi energy. $\mu_i$ depends on the experimental growth condition, which can be either N or Ga rich[89]. The atomic chemical potentials we use in our calculations are discussed in the Supplemental Material.

For adiabatic processes involving charged defects, to account for the polarisation in the outer MM region that is kept fixed in our simulations (see Fig, 1), we use Jost's formula[58]:
$$E_{pol} = -\frac{Q^2}{2R}(1 - \frac{1}{\varepsilon^0})$$
where $E_{pol}$ is the polarisation energy, $Q$ is the charge of the defect in the QM region, $R$ is the radius of the active region (i.e. where relaxations are treated explicitly) and $\varepsilon^0$ is the static dielectric constant. For vertical processes, however, where the charge state changes from $Q$ to $Q+\Delta Q$ but only electronic relaxation can occur, ionic polarisation due to the charge $Q$ will remain 'frozen in' at the final state, $Q+\Delta Q$. The generalised Jost's formula then takes the form:
$$E_{pol}^{vertical} = -\frac{(Q+\Delta Q)^2}{2R}\left(1 - \frac{1}{\varepsilon^\infty}\right) + \frac{Q^2}{2R}\left(\frac{1}{\varepsilon^\infty} - \frac{1}{\varepsilon^0}\right)$$
where $E_{pol}^{vertical}$ is the vertical polarisation energy, and $\varepsilon^\infty$ is the high-frequency dielectric constant. This expression is an update of that which we have used in previous works [58,62,90], hence there are minor differences in energy for some optical processes that we report here in comparison with those presented previously. For the values of the dielectric constants ($\varepsilon^0 = 10.09$ and $\varepsilon^\infty = 5.37$, see the Supplemental Material) we use one third of the trace of the corresponding dielectric tensors, calculated using the interatomic potentials[61,63].

Polarisation energies in dielectric materials by charged centres in the QM/MM methodology includes contributions from the QM, active MM and frozen MM regions, and Josts's formulae can provide a useful estimate of the two MM contributions using ratios of their respective radii. We thus calculate up to *ca.* 30% of the MM defect relaxation energy (more than 3 eV, for example, for triply charged defect states) to be due to the long-range polarisation effects. A similar conclusion can be drawn for periodic models, where long-range polarisation is usually not considered, but may prove to be vitally important in the total balance of defect energies.

## 3. Results

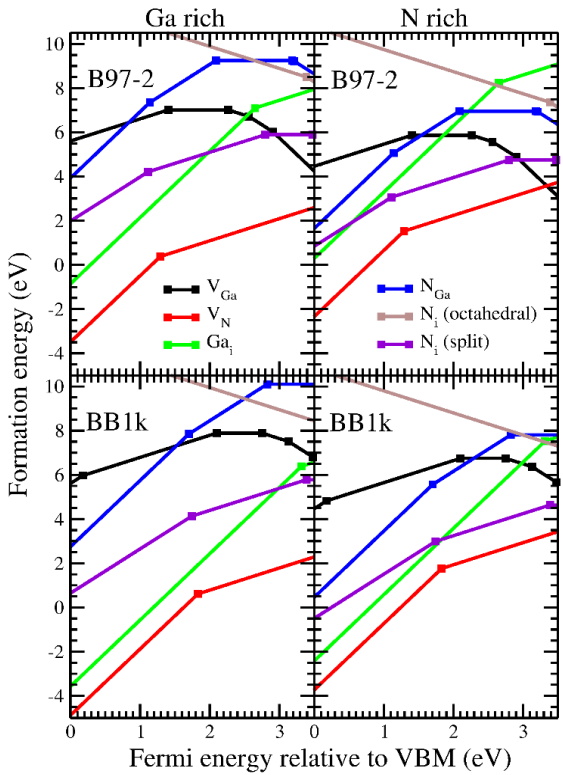

**Figure 2.** Point defect formation energies as a function of Fermi energy relative to the valence band maximum (VBM) of native defects in GaN determined using two functionals and under Ga-rich or N-rich conditions. The slopes of the lines indicate different charge states.

A summary of all native defects including their atomic and electronic structure along with vertical levels is given in the Supplemental Material while here we focus on the energetics of these defects and consequent observed physical properties.

### 3.1 Formation energies

We first consider the thermodynamic stability and adiabatic transitions of the point defects we consider in this work, *viz.* N and Ga vacancies, $V_N$ and $V_{Ga}$, interstitials, $N_i$ and $Ga_i$, and N antisite, $N_{Ga}$. (Based on the Madelung potential and normal oxidation states of Ga, we do not consider the Ga antisite as a plausible point defect in GaN across the whole range of relevant chemical potentials.) Our results are presented in figure 2. Note that the relative stability of a particular defect or charge state does not preclude formation by kinetic means, which could be observed experimentally.

We find, using either the B97-2 or BB1k XC density functional, that $V_N$ is the most favourable native point defect for Fermi energies within the band gap, apart from a small range close to the conduction band under N-rich conditions with B97-2. It is a donor-like defect throughout the band gap, stabilising in the +3 charge state above the valence band with a transition to the +1 state in the middle part of the gap. The +2 state of $V_N$ has high formation energy and is not favourable at any value of Fermi energy, indicating a negative U nature of the defect[91], which is analogous to isoelectronic F centres in many oxide materials as discussed in section 4 in Supplemental Material. The formation energy of $V_N$ in the +3 state near the valence band is negative, implying a spontaneous formation that would compensate any p-type carriers if the Fermi energy were close to the VBM. This behaviour fundamentally precludes any macroscopic samples from being p-type under thermodynamic equilibrium as any acceptors will be compensated by the vacancies; we discuss this point further in the next section. $V_N$ undergoes the (+3|+) transition at 1.29 eV (B97-2) and 1.83 eV (BB1k) above the VBM. Ganchenkova and Nieminen[32] reported that the -1 and -3 states of $V_N$ stabilised in the band gap below the CBM, in contrast to our calculations that show the stabilisation of the -1 charge state 0.32 eV (0.88 eV) above the CBM, determined using the B97-2 (BB1k) functional. Open-shell charged states 0 and -2, like the state +2, prove to be thermodynamically unstable under all physical conditions.

Under N-rich conditions near the conduction band, per the B97-2 calculations, $V_{Ga}^{3-}$, the only triple acceptor among the native defects in GaN, becomes more thermodynamically stable than $V_N$. This defect is a donor in the +1 charge state near the VBM, and it stabilises as an acceptor in the band gap in the charge states 0, -1, -2 and -3. With both functionals, near the VBM, $V_{Ga}$ is determined to be highly unstable.

Like the $V_N$, the $Ga_i$ is most favourable in the +3 charge state above the VBM and in the +1 state below the CBM, thus it remains a donor across the band gap. $Ga_i$ undergoes a (+3|+) transition in the upper half of the band gap, at 2.65 eV (B97-2) and 3.32 eV (BB1k) above the valence band. The formation

energy of $Ga_i^{3+}$ near the VBM is negative (except under N-rich conditions with the B97-2 functional, where the value is slightly positive, *ca*. 0.25 eV), implying a spontaneous formation, compensating positive free charge carriers, similar to the case of $V_N^{3+}$.

The formation of N interstitial defects can be considered as a competition between available octahedral and split interstitial sites. The octahedral site proves to be stable in the -1 state for BB1k across the band gap, while with B97-2 it undergoes a (-1|-2) charge transition at 0.11 eV below the CBM. This configuration, however, is unstable relative to the split interstitial for Fermi energies up to 1.37 eV (with B97-2) and 1.86 eV (with BB1k) above the CBM. On transition from the octahedral to the split geometry, the structure autoionizes, with an energy gain in excess of 2 eV. The B97-2 calculations show a (-1|0) thermodynamic transition at 0.2 eV below (with B97-2) and 0.28 eV above (with BB1k) the CBM. On lowering the Fermi energy into the band gap, the cationic charge states become more favourable – see figure 2.

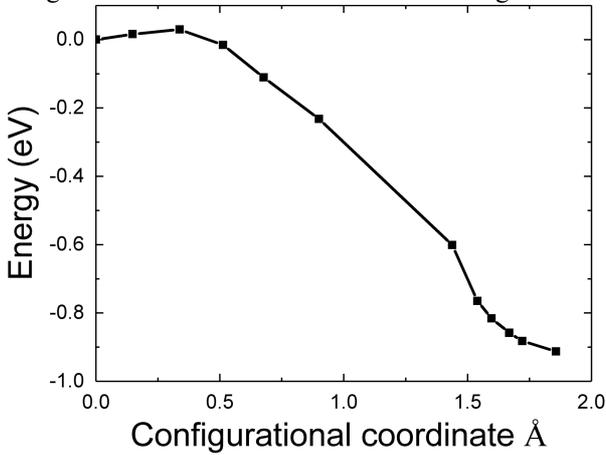

**Figure 3.** Transition state (climbing nudged-elastic band) calculation of $N_i$ from octahedral (left) to split (right) configurations in the charge state of -2, using the B97-2 density functional.

We employ the climbing-image nudged-elastic-band method to investigate the transition from an octahedral geometry to a more favourable split interstitial configuration in the -2 charge state; the result is shown in figure 3. Note that this transition is spin allowed (both configurations are of the same spin ½), whereas in the thermodynamically stable charge state of -1, the octahedral site is in a quasi-atomic state of $^3P_2$, while the ground state of the split interstitial is $^1\Sigma_g$, i.e the transition is spin forbidden. We determine an energy barrier of 0.03 eV, which is comparable to the thermal energy at room temperature. Therefore, we expect a transition from the octahedral to the split interstitial configuration to occur spontaneously under normal conditions.

Antisites include nitrogen-on-gallium ($N_{Ga}$) and gallium-on-nitrogen ($Ga_N$). $Ga_N$ will not be stable because of: (i) the large difference in the relative effective sizes of gallium and nitrogen ions; and (ii) the electron-poor valence shell of Ga (i.e. the electropositive nature of the metal Ga), which is highly unlikely to support Ga anionic states even in the presence of a favourable Madelung potential. Nitrogen, in contrast, can act both as an anion and cation in different chemical environments. $N_{Ga}$ is a donor-like defect near the VBM as it stabilises in the +3 charge state, but as the Fermi energy increases transitions occur first to the +2 state at 1.15 eV (B97-2) and 1.71 eV (BB1k) above the VBM, then to the neutral state at 2.09 eV (B97-2) and 2.83 eV (BB1k) above the VBM. Increasing the electron concentration further, using B97-2 we find that $N_{Ga}$ stabilises in the -1 state below the CBM, while it remains neutral when using the BB1k functional.

When comparing our results with previous studies for charged defects (see table 1), we find lower (higher) formation energies for positive (negative) charges and hence transition levels closer to the CBM and, equivalently, the vacuum reference energy. Therefore, in some cases, we disagree with previous work on predicted lowest energy charge states. We note, however, the calculated formation energies of neutral native defects are very close to those reported by other groups, as shown in tables 2 and 3. Gillen and Robertson[42] report that, for $V_{Ga}$, the stronger exchange interactions in the self-interaction-corrected sX-LDA approach result in defect levels higher in the band gap relative to the VBM, compared with standard LDA calculations. Lyons *et al.* calculated thermodynamic transition levels that are even higher above the VBM than those determined from sx-LDA calculations, and they link this behaviour to hole localization[44]. They emphasised the importance of full geometry optimisation with a hybrid functional and, as a demonstrative example, showed that hole localisation on nearest-neighbour N ions to a $V_{Ga}$ is accompanied by significant relaxation of those N ions[44], which we also observe in our calculations (see table S4). The trend for higher in-gap levels is confirmed by our hybrid DFT calculations, and, for the $V_{Ga}$, is even more pronounced than that of Lyons *et al.* who used the HSE functional. Indeed, we observe such a trend for all defects studied, with a stronger effect resulting from using the BB1k functional rather than the B97-2 functional. This observation is consistent with the more localised spin densities that result from BB1k functionals, as discussed in Supplemental Material.

**Table 1.** Thermodynamic transition levels of the $V_{Ga}$ with respect to the VBM. Label ~ indicates values estimated from figures in references. Note the trend of energy levels moving towards the CBM. The amount of the exact exchange in the functional is given in parentheses.

| $\epsilon$(eV) | LDA[23] | sX-LDA[42] | HSE[38] (28%) | HSE[44] (31%) | B97-2 (21%) | BB1k (42%) |
|---|---|---|---|---|---|---|
| +1\|0 | - | - | ~0.8 | 0.97 | 1.41 | 2.10 |
| 0\|-1 | 0.25 | 1.37 | ~1.7 | 1.68 | 2.27 | 2.75 |
| -1\|-2 | 0.7 | 1.88 | ~2.1 | 2.33 | 2.56 | 3.13 |
| -2\|-3 | 1.1 | 2.09 | ~2.4 | 2.80 | 2.91 | 3.48 |

**Table 2.** Defect formation energies in the neutral state in N-poor conditions. Label ~ indicates values estimated from figures in references. Method used in Ref: ab initio MD[16], LDA[17][23][28][92], PBE[25], HSE (28%)[38], LSDA[33], HSE (31%)[41], sX-LDA[42], GGA[93] (PW91).

| $E_f$(eV) | B97-2 | BB1k | LDA | HSE | PBE | MD, GGA |
|---|---|---|---|---|---|---|
| $V_N$ | 2.94 | 3.20 | ~3.6[42], 3.16[33] | ~3.3[38], ~3.1[41] | 2.59[25] | 3.2[16] |
| $V_{Ga}$ | 7.01 | 7.89 | ~7.1[42], 9.06[23], 8.40[33], 8.5[92] | ~7.9[38], ~7.8[26] | 7.02[25] | 8.2[16] |
| $Ga_i$ | 9.03 | 8.22 | | - | - | - |
| $N_i$ | 5.89 | 5.78 | 6.31[23] | ~6[38], ~5.9[26] | 6.19[25] | 5.73[93] |
| $N_{Ga}$ | 9.24 | 10.10 | ~7.8[23] | ~11[26] | 7.91[25] | - |

**Table 3.** Defect formation energies in the neutral state in N-rich conditions. Label ~ indicates values estimated from figures in references.

| $E_f$(eV) | B97-2 | BB1k | LDA | HSE | PBE |
|---|---|---|---|---|---|
| $V_N$ | 4.08 | 4.34 | 4.81[28] | ~4.2[38], ~4.5[41] | 3.03[25] |
| $V_{Ga}$ | 5.87 | 6.74 | ~6.5[17], 6.14[28], ~6.4[33] | ~6.6[38], ~6.6[26] | 6.58[25] |
| $Ga_i$ | 10.18 | 9.37 | 10.56[28] | - | - |
| $N_i$ | 4.75 | 4.63 | ~6.5[17], 4.57[28] | ~4.8[38], ~4.7[26] | 5.75[25] |
| $N_{Ga}$ | 6.95 | 7.80 | ~5.5[17], 5.21[28] | ~8.5[26] | 7.04[25] |

## 3.2 Defect and carrier concentrations

From the computed formation energies presented in the previous section, we can calculate the equilibrium concentrations and self-consistent Fermi energy, $E_F$ at a given temperature $T$, under the constraints of thermodynamic equilibrium and charge neutrality, using our program SC-FERMI[94,95]. The program works by calculating the concentration of each defect X in each charge state q from $[X]^q = N_X g_q \exp(-E_f^q/kT)$, where $N_X$ is the density of sites in which the defect may form, $g_q$ is the degeneracy of the state q (derived from the analysis in the proceeding section) and k is Boltzmann's constant, as well as determining the electron ($n_0$) and hole ($p_0$) concentrations by integrating the density of states weighted by the appropriate Fermi-Dirac function. Because the $[X]^q$ are functions of the Fermi energy (via the formation energy), as are $n_0$ and $p_0$, it is possible to determine the 'self-consistent' $E_F$ that satisfies the charge neutrality condition. The total concentration of each defect is obtained by summing the concentrations of the electronically stable charge states. With this procedure, fixed-concentration defects can be introduced and $E_F$ re-computed, thus modelling 'frozen in' defects species which may form and persist, e.g. due to kinetic barriers.

Under equilibrium conditions (see figure 4), we find that $V_N$ is the most dominant defect, but that the concentrations are low, as one would expect from the formation energies (figure 2). Using either XC functional results in very similar calculated concentrations and $E_F$. For N-poor conditions, $[V_N] > 10^{13}$ cm$^{-3}$ at $T$ ca. 800 K, while $n_0 = [V_N]$ for the temperature range studied ($0 < T < 1500$ K). $E_F$ remains in the band gap, from ca. 1.3 to 0.5 eV below the CBM, moving closer to the CBM as the temperature is increased. In N-rich conditions, due to the higher formation energies, $n_0$ and $[V_N]$ are several orders of magnitude below those of the N-poor case, and at higher $T$ (above 1200 K), minority carrier concentrations ($p_0$) of one (for B97-2) and 2 (for BB1k) orders of magnitude below that of the majority carriers are obtained. $E_F$ remains closer to the middle of the gap, varying from ca. 1.7 to 1.2 eV below the CBM as $T$ is increased up to 1500 K. These results demonstrate the intrinsic $n$-type nature of GaN, although the calculated carrier concentrations are low (below $10^{17}$ cm$^{-3}$).

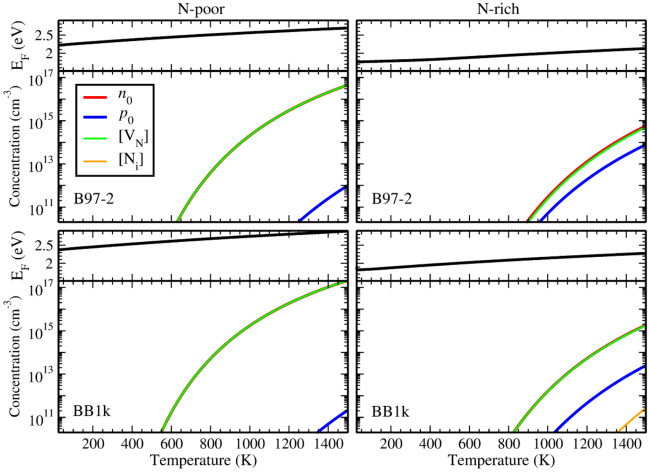

**Figure 4.** The calculated self-consistent Fermi energy ($E_F$, relative to the VBM, black line) and equilibrium concentrations of electrons ($n_0$, red line), holes ($p_0$, blue line), N split interstitials ([$N_i$], orange line) and nitrogen vacancies ([$V_N$], green line) as a function of temperature. Results from both XC functionals are shown for N-poor and N-rich conditions.

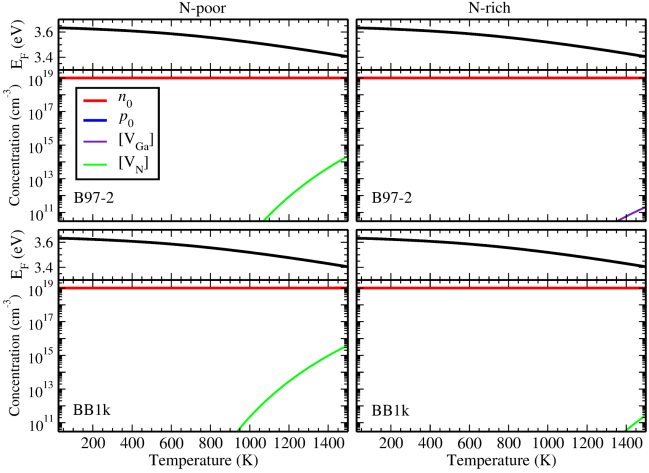

**Figure 5.** The calculated self-consistent Fermi energy ($E_F$, relative to the VBM, black line) and equilibrium concentrations of electrons ($n_0$, red line), holes ($p_0$, blue line), gallium vacancies ([$V_{Ga}$], indigo line) and nitrogen vacancies ([$V_N$], green line) as a function of temperature when a fixed concentration of fully ionised donors of $10^{19}$ cm$^{-3}$ is present. Results from both XC functionals are shown for N-poor and N-rich conditions.

Our procedure also allows us to analyse compensation of ionised donors or acceptors, by introducing a fixed concentration of either and determining the resulting defect concentrations and Fermi energy given the charge neutrality constraint. Turning first to donors: in figure 5 we show the calculated $E_F$ and defect and carrier concentrations when a fixed ionised donor concentration [$D^+$] = $10^{19}$ cm$^{-3}$ is present (D here could represent for example a Si$_{Ga}$ impurity, assumed to be fully ionised). Because of the lack of low formation energy negatively charged defects (see figure 2), even close to the CBM, [$D^+$] is uncompensated and $n_0 = 10^{19}$ cm$^{-3}$ over the full range of $T$ studied. In N-poor conditions, we do see a small concentration of $V_N$ at high $T$, which is several orders of magnitude lower than [$D^+$]. $E_F$ decreases as $T$ increases, following the curve expected for an ideal gas of fermions. In N-rich conditions, with the B97-2 functional, the formation energy of $V_{Ga}$ is below that of $V_N$ close to the CBM, but the resulting concentration of the order of $10^{12}$ cm$^{-3}$ is just noticeable in figure 5. If we vary [$D^+$], we only see deviation from $n_0 = $ [$D^+$] below a cut-off concentration [$D^+$] = $10^{15}$ cm$^{-3}$ (for $T$ = 900 K). Below this cut-off, $n_0 = $ [$V_N$]. The cut-off increases at higher $T$, but decreases going from N-poor to N-rich conditions.

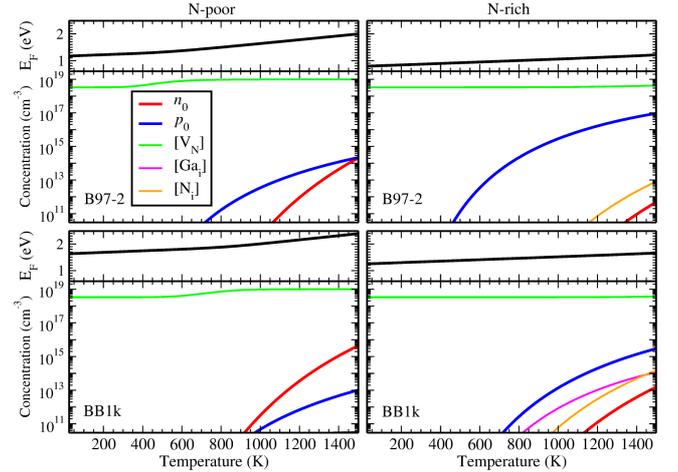

**Figure 6.** The calculated self-consistent Fermi energy ($E_F$, relative to the VBM, black line) and equilibrium concentrations of electrons ($n_0$, red line), holes ($p_0$, blue line), nitrogen vacancies ([$V_N$], green line), gallium interstitials ([$Ga_i$], magenta line) and N split interstitials ([$N_i$], orange line) as a function of temperature when a fixed concentration of fully ionised acceptors of $10^{19}$ cm$^{-3}$ is present. Results conditions.

For acceptors, the situation is quite different. Shown in figure 6 are $E_F$ and the defect and carrier concentrations when a fixed ionised acceptor concentration of $[A^-] = 10^{19}$ cm$^{-3}$ is present (here A could represent e.g. an ionised Mg$_{Ga}$). In this case, the acceptor is compensated by positive V$_N$, rather than holes, while $E_F$ remains within the band gap, at least 0.9 eV above the VBM, and moves further from the VBM as $T$ increases. Under N-rich conditions, using either XC functional results in $E_F$ that varies between 0.9 and 1.7 eV, with compensation by $[V_N^{3+}] = \frac{1}{3}[A^-]$. At $T > 700$ K, thermal excitation of holes results in $p_0 > 10^{13}$ cm$^{-3}$, which is still several orders of magnitude below that of $[A^-]$. Using the BB1k functional, we find that concentrations of other positively charged defects (Ga$_i$ and N$_i$) become greater than $10^{13}$ cm$^{-3}$ at $T > 1300$ K. Under N-poor conditions, the results are similar; but in this case, due to the lower V$_N$ formation energy, $E_F$ remains closer to the middle of the gap. As a consequence of this position of $E_F$, at $T > 400$ K (600 K) for B97-2 (BB1k) the lowest energy state of V$_N$ changes from 3+ to 1+, resulting in an increased compensation concentration (i.e. the compensation $[V_N^{3+}] = \frac{1}{3}[A^-]$ changes to $[V_N^+] = [A^-]$). At elevated temperatures, thermal excitation of carriers occurs, but as $E_F$ is closer to midgap both carrier types are observed.

From these results, we can deduce how point defect compensation further restricts effective acceptor activation. We focus on the N-rich conditions, where V$_N$ are less favourable to form and, therefore, one would expect acceptors to be compensated less. We have assumed that the acceptors are fully ionised, but the best-case scenario for p-type activity is at $T = 1500$ K for B97-2, where $p_0 = 10^{-2}$ $[A^-]$. For Mg impurities, many studies claim that the ionisation energy is ~0.2 eV[15,96]. Our results show that, even at 1500 K, where one would expect $p_0$ of the order of 0.1 [Mg$_{Ga}^-$] from the acceptor ionisation energy, the actual $p_0$ would be at most 10$^{-3}$ [Mg$_{Ga}^-$]. At lower $T$, $p_0$ decreases to many orders of magnitude lower than that of the acceptor concentration. Therefore, in order to see any significant p-type activity, the compensating V$_N$ must somehow be controlled, most likely through forming complexes and/or more extended defects, which would explain why no p-type activity is observed unless either the Mg concentration is very high (at least $10^{19} – 10^{20}$ cm$^{-3}$), or particular structuring and design is employed (e.g. thin films[97], capping layers[78]). Under N-rich conditions, that help to suppress N vacancy formation, keeping the V$_N$ concentration one order of magnitude below the acceptor concentration will result in a 70% activation of the acceptors at room temperature (in our example, $p_0 = 7 \times 10^{18}$ cm$^{-3}$). Furthermore, we find that, at elevated temperatures, increasing $[A^-]$ results in further compensation by other positive defects, particularly $[N_i^{2+}]$, but also $[Ga_i^{3+}]$ when using the BB1k functional.

One of the puzzles in the defect chemistry of GaN is the apparent contradiction between the high formation energies for Ga vacancies calculated consistently by different groups[23,25,27,28,32,42,43,92] and several reports of substantial concentrations, up to $10^{20}$ cm$^{-3}$, of this defect as seen in positron annihilation spectroscopic experiments[98–101]. Our calculations, as can be seen from figure 4, do not confirm the presence of such concentrations under any conditions in a pristine (formally undoped) material. A resolution for this contradiction has been presented by Saarinen *et al.* who have shown that the Ga vacancies in GaN compensate for extrinsic donors such as O impurities and their concentration drops below $10^{15}$ cm$^{-3}$ when [O] is reduced[100]. Moreover, V$_{Ga}$-O$_N$ defect complexes have been proposed as the main form in which the vacancy is accommodated by the material[102]. Modelling defect complexes is outside the scope of the current study, but we do expect a significant stabilization of both the vacancy and impurity by the Coulomb interaction between them as nearest or next nearest neighbours, which would result in a substantial rise in the observed vacancy concentration.

*3.3 Ionization energies as defect (transition) levels*

The knowledge of defect formation energies in different oxidation (or charge) states – *cf.* reference [103] – alongside the vertical ionisation potentials and electron affinities provides insight into many defect processes that play a key rôle in photo-absorption and luminescence, charge trapping, *etc.* In this analysis, we use a conventional configurational coordinate diagram (see reference [104]) shown in figure 7. For example, an electron trapped by an acceptor in the ground state (A$^-$) can be excited to the conduction band if it absorbs a photon with sufficient energy ($E_{ab}$), the resulting configuration being the neutral charge state plus a conduction electron, A$^0$ + e$^-$. The defect in the ionised state will relax ($E_{rel}$*) before emitting a photon and returning to its ground state via an electron-hole recombination. The relaxation lowers the energy of the ionised state as the surrounding atoms move to new stable positions. On emission, a conduction electron undergoes a vertical transition from the conduction band to the empty defect state (A$^-$), i.e. recombining with the hole bound to the acceptor, giving rise to a photoluminescence ($E_{PL}$). Again, the atoms around the defect relax back to the initial ground state ($E_{rel}$) after the optical emission. An analogous process will involve a hole ionisation from an acceptor, which, in other words, describes a capture by the acceptor of an electron from the valence band – see the upper part of the configuration-coordinate diagram shown in figure 7. The energy difference between the minima of the ionised and ground states can be identified in the photoluminescence spectra as a zero-phonon line (ZPL). In the adiabatic approximation following the Franck-Condon principle, the vertical electron transition is assumed to be much faster than the atomic relaxation, which is usually the case in defect processes[15], although we note that other more complex processes involving other photo-generated carriers may also be relevant[74]. Furthermore, the total energies of both the ground and excited electronic states are shifted up by the corresponding phonon contributions, which results in the next order corrections to the transition energies. The defect-to-band transitions are also often associated with activation of charge carriers trapped at defect sites, so the ZPL energies are considered as activation energies in the relevant electrical processes, which is the topic of the next section.

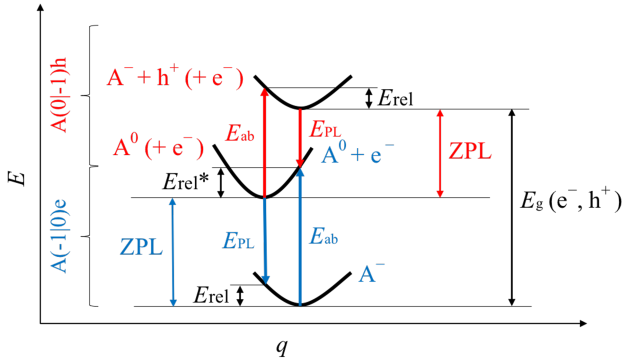

**Figure 7.** A configuration-coordinate ($q$) diagram of ionisation and capture by an acceptor A of an electron, e$^-$ shown in blue – A(−1|0)e – and of a hole, h$^+$ shown in red – A(0|−1)h. Electrons and holes are assumed to be at conduction and valence band edges, respectively; separated by the band gap ($E_g$). Zero phonon lines (ZPL) characterise thermodynamic (adiabatic) transitions; and photo-absorption and photoluminescence energies ($E_{ab}$ and $E_{PL}$ respectively) are parameters of optical (vertical) processes. Defect atomic relaxation energies, $E_{rel}$ and $E_{rel*}$, are interpreted as Stokes shifts.

The relationship with the defect level diagrams discussed in section 4 in the Supplemental Material is as follows: the position of the donor (occupied) level of a defect in charge state Q with respect to the conduction band corresponds to the photo-absorption in a D(Q|Q+1)e process, while its separation from the valence band gives the photoluminescence energy in a D(Q|Q−1)h process; the corresponding acceptor levels yield the energies of photo-absorption in a A(Q|Q−1)h process and that of photoluminescence in a A(Q|Q+1)e process. Further calculations of the potential energy surface along the configuration coordinate could be used to derive line shapes and intensities, but such calculations are beyond the scope of the current work.

We now can compare our results, presented in table 4, with pertinent available experimental data from spectroscopic measurements. We focus on the values computed using the BB1k functional, which is expected to provide a better account of electron localisation and, consequently, more accurate ionisation energies. The B97-2 derived results are provided in table 4 for comparison.

The observed photoluminescence spectra from unintentionally and intentionally doped *n*-type GaN almost always contain a broad yellow luminescence (YL) band with a maximum at about 2.2 eV[15]. Originally, the yellow band has been assigned to a radiative transition from a shallow donor or from the conduction band to a deep acceptor[105,106]. Previously, V$_{Ga}$ as a native defect has been suggested to be involved in YL both on computational[42,43] and experimental[98,101] evidence. In our calculations (see table 4), the negative and neutral states of V$_{Ga}$ are non-radiative electron emission centres, and its +1 state can give rise to an emission peak for radiative electron capture in the infrared region, V$_{Ga}$(0|+1)e, which we discuss in more detail below. More positively charged states of the defect prove to be unstable and we conclude that the original assignment is not supported by our data. It has been, however, reported that the observed increase in intensity of the 2.2 eV band results from redistribution of holes released on excitation to the valence band[107]. Indeed, we find an emission peak for radiative hole capture by the −1 and −2 states of V$_{Ga}$ at 2.17 and 2.21 eV; this process, however, will only be one of the sources of the widely observed YL. The respective calculated ZPL energies of 2.75 and 3.13 eV for both processes are for example significantly higher than the recently reported value of 2.57 eV for the YL1 centre[108]. We note that we calculate consistently high (> 0.5 eV) defect relaxation energies, which is not consistent with such a low ZPL energy (by implication $E_{rel}$ ~ 0.37 eV). We find only one other defect state with the PL in the YL range: an octahedrally coordinated interstitial N (N$_i^O$) in the −1 state would produce an emission peak at 2.17 eV on radiative hole capture. The octahedral interstitial configuration with a triplet electronic ground state is though less stable than the singlet split interstitial structure. If formed it would be a relatively long-lived species, for example, in materials under ionising irradiation with energy above that of the N displacement threshold or grown under nitrogen rich conditions, and GaN thin films or powders exposed to nitrogen under pressure. We speculate that other sources of YL relate to defect complexes and/or impurities, which we have not considered in the present study.

Yan et al.[41] have claimed that, in p-type GaN, $V_N$ in the +3 state can make a transition to the +2 state and give rise to an emission peak at 2.18 eV, which is in the yellow region of the spectrum. We do not observe such a peak for $V_N$.

Another sharp peak at about 3.25-3.27 eV is observed in undoped GaN at low temperatures, which is known as the shallow DAP (donor to acceptor pair) band and is caused by transitions from the shallow donors to the shallow acceptors[15,109–111]. Employing the present analysis, we do not recognize this peak in our calculations – see however further discussion in section 3.5.

In GaN layers grown by metal-organic chemical vapour deposition (MOCVD) or hydride vapour phase epitaxy (HVPE), a blue luminescence (BL) band is often observed in the low-temperature PL spectrum at about 2.9 eV with a 3.098 eV ZPL[107]. It has been attributed to transitions from the conduction band or a shallow donor to a relatively deep acceptor having an ionization energy of about 0.34-0.4 eV[15]. We calculate an emission peak for radiative electron capture by the -3 and -2 state of octahedral $N_i$ at 2.88 and 2.76 eV, respectively, but this configuration is significantly less stable than the split $N_i$, whereas the ZPL energies are much higher than the observed value. Another peak at 2.85 eV for radiative electron capture by the -2 state of $N_{Ga}$ is found; however, this charge state remains metastable across the band gap.

In undoped GaN grown by HVPE or MOCVD methods, a red luminescence (RL) band is sometimes observed at about 1.8 eV, and is attributed to a deep acceptor level at 1.130 eV above the VBM[111,112]. We determine an emission peak for radiative hole capture by the neutral state of $N_{Ga}$ at 1.80 eV, while the +1 state of $N_{Ga}$ is calculated at 1.15 eV above the VBM ($E_{abs}$ for $N_{Ga}(+1|+2)e$ is 2.35 eV, see figure S6). Of course, given the high computed formation energy of the $N_{Ga}$, these peaks would only be observed in samples prepared by non-equilibrium processes. Katsikini et al. find a RL1 appearing 1.7 eV below the absorption edge that is attributed to N interstitials – which we cannot confirm – and a RL2 appearing at about 1.0 eV above the absorption edge which is attributed to nitrogen dangling bonds[113]. We find, however, that the +2 and +1 state of $Ga_i$ would give rise to emission peaks for hole capture at 1.89 eV and 1.83 eV. As $Ga_i^+$ would have the highest concentration of these defects in n-type Ga-rich conditions, we expect that this defect is a significant source of red PL in the corresponding samples of GaN. In N-rich conditions however, the concentration of $Ga_i$ approaches that of $N_{Ga}$ and, perhaps, both defects would compete for recombination with the photogenerated holes, although of course the two defects could form a complex with different spectroscopic signatures.

Next, a broad green band (GL) at about 2.5 eV and a broad red band at 1.9 eV have been detected in highly resistive GaN samples grown by MBE under extremely Ga-rich conditions, and have provisionally been attributed to internal transitions in some defects[15,114]. We find an emission peak for radiative hole capture $N_{Ga}^0$ at 1.80 eV, and another emission peak for radiative hole capture by $N_{Ga}^-$ at 2.54 eV, both with large relaxation energies that would account for the broad nature of the PL and which could provide a reasonable alternative attribution – see figure 8. Note that the $N_{Ga}^0$ level lies in the middle of the band gap.

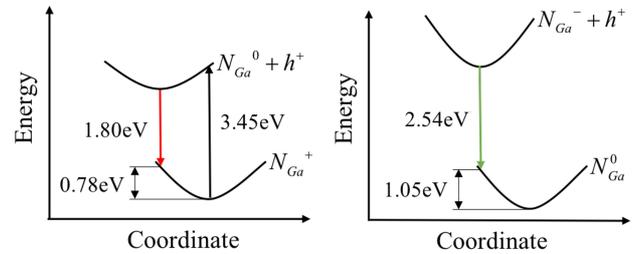

**Figure 8.** Configuration-coordinate diagrams for $N_{Ga}$ using the BB1k density functional.

Furthermore, the GL bands will be originating from other intrinsic defects as, for example, a split $N_i^0$ has a green emission peak for radiative hole capture at 2.49 eV, and a $V_{Ga}^{-3}$ has a green emission peak for radiative hole capture at 2.35 eV, which should then be expected to be seen in GaN prepared in N-rich conditions.

As seen from table 4, we find that a $V_N^0$ and $V_N^-$, on a hole capture from the VBM, are likely to give rise to the PL peaks at 3.46 eV and 3.41 eV that are practically ubiquitously observed in GaN. We have identified the $V_N$ as the major compensating centre for acceptor impurities in GaN (see section 3.2 above), where hole capture by electron-rich shallow donors explains the near band edge emission (see refs. [62] and [74]). We discuss the 3.46 eV emission peak in further detail in section 3.5.

Finally, a near infrared (IR) PL signal at 0.95 eV (see refs. [115–118]) appears in electron irradiated samples; this PL band mainly vanishes on annealing. Our calculations allow us to ascribe this band to electron capture by a $V_{Ga}^+$ (see figure 9). The $V_{Ga}$ defects have previously been proposed as the obvious result of the irradiation and a convenient position of the 1.1 eV hole capture transition by $V_{Ga}^{-3}$ from early DFT calculations. The PL has been shown to be correlated with magnetically detected L5 and L6 defect centres in GaN, which in turn were associated with $Ga_i$, situated at differing positions with respect to $V_{Ga}$. This band is anticorrelated to the RL observed before irradiation. Above, we have attributed the RL to either $Ga_i$ or $N_{Ga}$. Electron irradiation should remove $N_{Ga}$ and result in a significant concentration of $V_{Ga}$, which will bind electrons on nitrogen ions decorating the vacant site (reminiscent of the RL2 centres as proposed by Katsikini *et al*.). On photoexcitation, significant hole concentrations will form at the vacancies (as the photoexciting frequency is sub-band gap), rather than in the delocalised band states, which will suppress the hole capture process involved in the $Ga_i$ RL, and promote the $V_{Ga}$ electron capture process described above. Therefore, we would expect the RL to reduce in intensity and the IR PL to increase on irradation, as is seen experimentally. Subsequent annealing will remove some of the vacancies, while those remaining will adopt the lowest energy charge state (−2 or −3 under *n*-type conditions, see figure 2), which cannot luminesce in the infrared, resulting in a suppression of IR PL, again as seen experimentally.

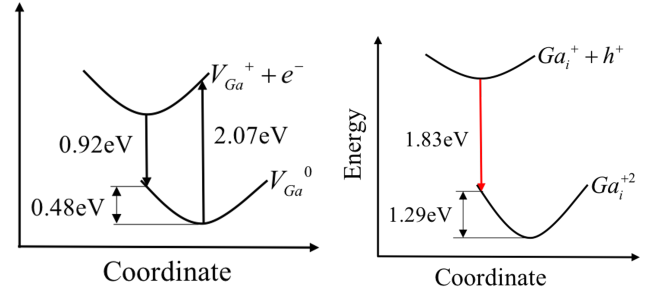

**Figure 9.** Configuration-coordinate diagrams for $V_{Ga}$ and $Ga_i$ using the BB1k density functional.

In this section, we have compared our results to a selection of experimental PL studies. We do not attempt to be exhaustive in our comparison with the available literature, or to discuss the spectroscopic processes in intricate detail. Such a study on the luminescence properties of undoped GaN is beyond the scope of the present work. Our results given in table 4 should, however, provide a useful comparison tool for the interested reader.

**Table 4.** Summary of calculated ionization energies for the full range of point defects and charge states. The energies terms are defined in figure 7, and all values are in eV. Results are given for two density functionals, but in general, we expect more accurate values when using the BB1k functional. Optical processes involving relaxed resonant states are not included here.

| Defect states | B97-2 | | | | | BB1k | | | | |
|---|---|---|---|---|---|---|---|---|---|---|
| | $E_{PL}$ | $E_{ab}$ | ZPL | $E_{rel}$ | $E_{rel}^*$ | $E_{PL}$ | $E_{ab}$ | ZPL | $E_{rel}$ | $E_{rel}^*$ |
| $V_N(+2\|3)e$ | 0.75 | 3.64 | 2.17 | 1.42 | 1.46 | 0.08 | 3.16 | 1.54 | 1.46 | 1.63 |
| $V_N(+1\|+2)e$ | 1.29 | 3.47 | 2.25 | 0.96 | 1.22 | 0.82 | 3.04 | 1.81 | 0.99 | 1.23 |
| $V_N(0\|+1)e$ | −1.03 | 0.59 | −0.35 | 0.68 | 0.94 | −1.68 | 0.04 | −0.91 | 0.77 | 0.95 |
| $V_N(-1\|0)e$ | −1.08 | 0.60 | −0.26 | 0.82 | 0.86 | −1.67 | 0.09 | −0.84 | 0.83 | 0.94 |
| $V_N(+3\|+2)h$ | −0.14 | 2.75 | 1.32 | 1.46 | 1.42 | 0.34 | 3.43 | 1.97 | 1.63 | 1.46 |
| $V_N(+2\|+1)h$ | 0.03 | 2.22 | 1.19 | 1.22 | 0.96 | 0.46 | 2.68 | 1.69 | 1.23 | 0.99 |
| $V_N(+1\|0)h$ | 2.91 | 4.53 | 3.85 | 0.94 | 0.68 | 3.46 | 5.19 | 4.41 | 0.95 | 0.77 |
| $V_N(0\|-1)h$ | 2.90 | 4.58 | 3.76 | 0.86 | 0.82 | 3.41 | 5.17 | 4.35 | 0.94 | 0.83 |
| $V_{Ga}(-3\|-2)e$ | −0.69 | 1.59 | 0.60 | 1.29 | 0.99 | −1.36 | 1.15 | 0.02 | 1.38 | 1.13 |
| $V_{Ga}(-2\|-1)e$ | 0.15 | 1.58 | 0.94 | 0.79 | 0.64 | −0.64 | 1.29 | 0.37 | 1.01 | 0.92 |
| $V_{Ga}(-1\|0)e$ | 0.61 | 1.73 | 1.23 | 0.62 | 0.49 | −0.15 | 1.33 | 0.75 | 0.90 | 0.58 |
| $V_{Ga}(0\|+1)e$ | 1.72 | 2.46 | 2.11 | 0.39 | 0.36 | 0.92 | 2.07 | 1.59 | 0.48 | 0.66 |
| $V_{Ga}(-2\|-3)h$ | 1.92 | 4.19 | 2.91 | 0.99 | 1.29 | 2.35 | 4.86 | 3.48 | 1.13 | 1.38 |
| $V_{Ga}(-1\|-2)h$ | 1.92 | 3.36 | 2.56 | 0.64 | 0.79 | 2.21 | 4.15 | 3.13 | 0.92 | 1.01 |
| $V_{Ga}(0\|-1)h$ | 1.77 | 2.89 | 2.26 | 0.49 | 0.62 | 2.17 | 3.65 | 2.75 | 0.58 | 0.90 |
| $V_{Ga}(+1\|0)h$ | 1.04 | 1.79 | 1.40 | 0.36 | 0.39 | 1.44 | 2.58 | 2.10 | 0.66 | 0.48 |
| $Ga_i(+2\|+3)e$ | −0.47 | 2.29 | 0.51 | 0.98 | 1.38 | −1.49 | 1.61 | −0.01 | 1.48 | 1.62 |
| $Ga_i(+1\|+2)e$ | −0.19 | 2.18 | 1.19 | 1.38 | 1.00 | −1.00 | 1.67 | 0.38 | 1.38 | 1.29 |
| Ga$_i$(0\|+1)e | −1.28 | −0.90 | −1.09 | 0.19 | 0.19 | −2.10 | −0.63 | -1.65 | 0.45 | 1.02 |
| Ga$_i$(+3\|+2)h | 1.21 | 3.97 | 2.59 | 1.38 | 0.98 | 1.89 | 4.99 | 3.51 | 1.62 | 1.48 |
| Ga$_i$(+2\|+1)h | 1.32 | 3.70 | 2.32 | 1.00 | 1.38 | 1.83 | 4.50 | 3.12 | 1.29 | 1.38 |
| Ga$_i$(+1\|0)h | 4.40 | 4.78 | 4.59 | 0.19 | 0.19 | 4.13 | 5.60 | 5.15 | 1.02 | 0.45 |
| N$_i^O$(−3\|−2)e | −1.98 | 1.09 | −0.34 | 1.64 | 1.44 | −2.72 | 0.63 | −0.92 | 1.80 | 1.54 |
| N$_i^O$(−2\|−1)e | −1.12 | 1.19 | 0.12 | 1.24 | 1.07 | −1.75 | 0.74 | −0.43 | 1.32 | 1.18 |
| N$_i^O$(−1\|0)e | - | 1.66 | - | - | - | - | 1.33 | - | - | - |
| N$_i^O$(−2\|−3)h | 2.41 | 5.49 | 3.85 | 1.44 | 1.64 | 2.88 | 6.23 | 4.42 | 1.54 | 1.80 |
| N$_i^O$(−1\|−2)h | 2.31 | 4.62 | 3.38 | 1.07 | 1.24 | 2.76 | 5.26 | 3.94 | 1.18 | 1.32 |
| N$_i^O$(0\|−1)h | 1.85 | - | - | - | - | 2.17 | - | - | - | - |
| N$_i^S$(−2\|−1)e | −1.99 | −1.35 | −1.60 | 0.39 | 0.25 | −2.85 | −2.17 | −2.45 | 0.40 | 0.28 |
| N$_i^S$(−1\|0)e | −0.77 | 1.28 | 0.03 | 0.80 | 1.25 | −0.51 | 0.84 | −0.31 | 0.20 | 1.15 |
| N$_i^S$(0\|+1)e | −0.15 | 1.68 | 0.70 | 0.85 | 0.97 | −0.80 | 1.02 | 0.12 | 0.92 | 0.90 |
| N$_i^S$(+1\|+2)e | 0.99 | 3.44 | 2.39 | 1.40 | 1.05 | 0.13 | 3.11 | 1.76 | 1.63 | 1.35 |
| N$_i^S$(+2\|+3)e | - | 4.12 | - | - | - | - | 3.70 | - | - | - |
| N$_i^S$(0\|−1)h | 2.23 | 4.28 | 3.48 | 1.25 | 0.80 | 2.66 | 4.01 | 3.81 | 1.15 | 0.20 |
| N$_i^S$(+1\|0)h | 1.83 | 3.65 | 2.80 | 0.97 | 0.85 | 2.49 | 4.30 | 3.39 | 0.90 | 0.92 |
| N$_i^S$(+2\|+1)h | 0.07 | 2.52 | 1.12 | 1.05 | 1.40 | 0.40 | 3.37 | 1.75 | 1.35 | 1.63 |
| N$_i^S$(+3\|+2)h | −0.62 | - | - | - | - | −0.20 | - | - | - | - |
| N$_{Ga}$(−2\|−1)e | −1.16 | 1.11 | 0.29 | 1.45 | 0.82 | −2.40 | 0.65 | −0.22 | 2.18 | 0.87 |
| N$_{Ga}$(−1\|0)e | −0.65 | 1.08 | 0.32 | 0.97 | 0.76 | −1.49 | 0.97 | −0.08 | 1.41 | 1.05 |
| N$_{Ga}$(0\|+1)e | 1.00 | 1.95 | 1.50 | 0.50 | 0.46 | 0.06 | 1.70 | 0.93 | 0.87 | 0.78 |
| N$_{Ga}$(+1\|+2)e | 0.59 | 2.90 | 1.34 | 0.75 | 1.48 | −0.13 | 2.35 | 0.43 | 0.56 | 1.91 |
| N$_{Ga}$(+2\|+3)e | 1.08 | 3.60 | 2.36 | 1.28 | 1.23 | 0.38 | 3.08 | 1.80 | 1.42 | 1.28 |
| N$_{Ga}$(−1\|−2)h | 2.40 | 4.66 | 3.22 | 0.82 | 1.45 | 2.85 | 5.90 | 3.72 | 0.87 | 2.18 |
| N$_{Ga}$(0\|−1)h | 2.42 | 4.16 | 3.18 | 0.76 | 0.97 | 2.54 | 5.00 | 3.59 | 1.05 | 1.41 |
| N$_{Ga}$(+1\|0)h | 1.55 | 2.51 | 2.01 | 0.46 | 0.50 | 1.80 | 3.45 | 2.58 | 0.78 | 0.87 |
| N$_{Ga}$(+2\|+1)h | 0.60 | 2.92 | 2.08 | 1.48 | 0.75 | 1.16 | 3.64 | 3.07 | 1.91 | 0.56 |
| N$_{Ga}$(+3\|+2)h | −0.09 | 2.42 | 1.14 | 1.23 | 1.28 | 0.43 | 3.13 | 1.71 | 1.28 | 1.42 |

*3.4 Defect as traps*

Defect transition levels and configurational diagrams could usefully be employed in considering diverse phenomena that are not limited to the optical processes. A large number of deep electron and hole traps (see e.g. [119] and [120]) have been reported using a host of modern electrical, dielectric and related techniques. In our analysis, we distinguish fast optical (vertical) and slow thermodynamic (adiabatic) processes. Interpretation of experiment, however, can become quite involved and, in most cases, will require a separate study. Depending on the relative thermodynamic stability of the terms corresponding to the initial and final states, we can recognise thermally or optically activated ionisations, followed by recombination accompanied by luminescence or radiationless processes. Below, we consider examples of attribution of experimentally reported trap levels based on our defect calculations employing the BB1K functional.

We calculate the energy of $V_N^0$ at $E_c - 0.04$ eV (see figure S1) in excellent agreement with a recent optical admittance spectroscopy studies of a 0.051 eV state[121] and its singly negative state at $E_c - 0.09$ eV that falls in the range of ET1[120]. These levels indicate the shallow donor nature of the vacancy, and were discussed in a previous publication[62]. Both states, however, are metastable (as seen from the negative ZPL energies), and the majority of defects are expected to be in the singly or triply positively charged states depending on the position of the $E_F$ – see our discussion of the corresponding defect concentrations in section 3.2 above. In this case the vertical ionisation potential is our estimate of the activation energy. A hole trapping by $V_N^+$ at 1.69 eV is also close to the 1.76 eV energy of the H5 trap[119], for which we, however, find more candidates, e.g. a $V_N(+3|+1)2h$ transition that lies at $E_V + 1.83$ eV.

A dominant electron trap in variously prepared GaN is the E3[119] (or ET10[120]) centre, with its level at $E_c - 0.56$ eV [122,123] and high hole capture cross sections; thus acting as an efficient recombination centre. The concentration of the centres with the optical ionization energy of *ca.* 1.3 eV was found to correlate directly with the intensity of the YL band. These measurements are consistent with our calculated optical ionisations levels of 1.15 and 1.29 eV for $V_{Ga}^{-3}$ and $V_{Ga}^{-2}$, respectively, with the thermal transition between the two lying about 0.02 eV below the $E_c$. Under the thermal activation, the latter defect is, therefore, expected to dominate, and the hole recombination on the defect centre, we predict, will result in an optical transition in the YL range (emission at 2.35 eV). As the PL energy in the electron ionisation case is calculated to be negative, the activation energy is ill defined, which will qualitatively change the kinetic behaviour of the system. The E3 level could be attributed to $V_{Ga}^{-2}$ or, indeed, any of the three negative states of $V_{Ga}$ if the activation energy lies somewhere between vertical and adiabatic levels. Provisionally, the activation energy for $V_{Ga}(-1|0)e$ would thus be expected to be between 0.75 and 0.90 eV, which would place this transition in the energy range of ET12 donor centres whose presence, however, is usually linked to dislocations[120].

The presence of a split interstitial nitrogen has been convincingly demonstrated by single and double electron spin resonance techniques[31], which is magnetically active in the neutral charge state, $N_i^{S0}$. We calculate its thermal ionisation level at $E_c-0.12$ eV, close to the activation energy of ET2 centre[120]. Its further thermal ionisation energy of 1.76 eV is not accessible by usual DLTS methods but we calculate a counterpart hole trapping in the $N_i^S(+2|+1)h$ process as 1.75 eV, which is again in a marked agreement (or coincidence) with the $E_V + 1.76$ eV position of the H5 trap[119].

For the interstitial Ga, the experimental evidence is again from the optically detected magnetic resonance[35], which is observed within a broad IR PL band around 0.95 eV. The IR PL band emerges on electron irradiation of GaN, and its magnetic structure has been attributed to the $Ga_i^{+2}$. The negative values of $E_{PL}$ we calculate for the neutral and positively charged states of $Ga_i$ are indicative of the radiationless character of the involved processes and the corresponding bounds on the activation energies are rather wide. The neutral state is predicted to be autoionized, and the thermal ionisation energy of $Ga_i^{+1}$ to be 0.38 eV. Perhaps, more usefully, the energy of the $Ga_i(+1|+3)2e$ transition is found to be 0.18 eV, which places this defect midway between ET3 and ET4[120].

In n-type GaN, the $N_{Ga}$ defect adopts a neutral charge state with the thermal ionisation potential of 0.93 eV, which makes it a primary candidate for the ET12 centre or, possibly, a slightly lower energy ET11[120,124], and which could also be related to the E8 traps as seen in ref [119]. A similar attribution has been proposed to a $E_c - 1.04$ eV trap with the thermal activation of 0.90 eV from *I–T* characterisations by Polenta *et al.*[125]. The activation energy of further ionisation $N_{Ga}(+1|+2)e$ can be estimated to lie between 0.43 and 0.56 eV. It is unlikely that this process is responsible for the ubiquitous $E_c - 0.56$ eV trap discussed above – we expect the antisite to be only a small minority defect, but it is consistent with the behaviour of rare ET8 traps at $E_c - 0.49$ eV[126], which have previously been associated with $N_{Ga}$, and show a barrier of about 0.15 eV for electron capture (see discussion in [120]). Finally, the hole capture by $N_{Ga}^{+3}$ at 1.71 eV is again consistent with the position of H5 traps[119], for which however we have already seen more candidates.

In conclusion we reiterate that identification of particular defect processes is not possible on the closeness of the calculated energies with the reported trap levels or activation energies, but will require further careful analysis.

*3.5 Effects of delocalization*

Analysis of the defect formation and ionisation energies presented above assumes a compact nature of the defect electronic states. Such an assumption, however, is only an approximation as in real dielectric materials any charged centre will necessarily trap one or more charge carriers in shallow hydrogenic states (cf. Ref. [57] & [127]), which are typically probed by DLTS and related techniques. Due to the highly delocalised character of such states the exact nature of the defect becomes less important and different defects of the same charge would have very similar energies. When comparing our results to photoluminescence and other experiments, we have followed the standard procedure of considering ionisation to and from bands. A more complete picture would involve alternative levels associated with the diffuse states, which, however, lies outside of the scope of this study. As a guide to the interested reader we provide the characteristic energies and effective Bohr radii of hydrogenic defect traps both for electrons ($m^*$=0.22 $m_e$ [from [57]]) and holes ($m^*$=1.50 $m_e$ [[128]]) in Table 5. As one can see, the electron states in GaN are actually diffuse, whereas holes remain quite compact.

We find that the diffuse hole level is 0.2 eV above the VBM, indicating that a shallow acceptor would bind a hole with this ionisation energy. The only means of achieving p-type GaN is through doping with high concentrations of Mg, and an associated level at ~0.2 eV is observed experimentally. An adiabatic ionisation energy of ~0.2 eV is commonly attributed to the (0/-1) transition of $Mg_{Ga}$ from previous DFT studies. The diffuse hole level, however, indicates that a much shallower acceptor that binds effective mass like holes must predominate in p-type GaN, given the agreement with the experimentally observed level. Moreover, as discussed above, we attribute the ubiquitous 3.46 eV UV PL peak seen in undoped and p-doped GaN to the recombination of an electron bound to a neutral $V_N$ with a hole at the VBM (see figure 10). Another peak observed experimentally in a broad range of samples, in particular p-type GaN, is DAP peak at 3.25-3.27 eV. If, instead of recombining with a hole at the VBM, the electron bound to the $V_N$ recombines with a diffuse hole bound to a shallow acceptor, the emission energy will shift by 0.2 eV, resulting in a PL peak at 3.26 eV. The balance between these two competing processes can explain the observed peaks, and demonstrates the power of including the diffuse states in analysis of spectroscopic processes.

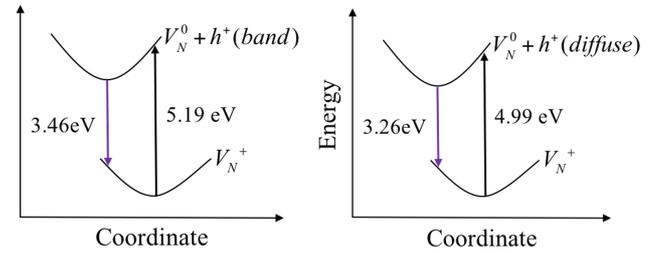

**Figure 10.** Configuration-coordinate diagrams for $V_N$ using the BB1k density functional.

**Table 5.** Energies and effective Bohr radii of hydrogenic defect traps in GaN for charge state $q$.

| Adiabatic | electrons | | holes | |
|---|---|---|---|---|
| $|q|$ | Radius (Å) | Energy to band edge (eV) | Radius (Å) | Energy to band edge (eV) |
| 1 | 24.270 | -0.029 | 3.560 | 0.200 |
| 2 | 12.135 | -0.059 | 1.780 | 0.401 |
| 3 | 8.090 | -0.088 | 1.187 | 0.601 |
| | | | | |
| Vertical | electrons | | holes | |
| $|q|$ | Radius (Å) | Energy to band edge (eV) | Radius (Å) | Energy to band edge (eV) |
| 1 | 12.917 | -0.104 | 1.894 | 0.708 |
| 2 | 6.458 | -0.208 | 0.947 | 1.415 |
| 3 | 4.306 | -0.311 | 0.631 | 2.123 |

## 4. Conclusion

We have reported a detailed systematic study of point defect processes in GaN from embedded cluster calculations. Quantitative energy levels of each native defect in GaN were provided. We find a shallow donor level of $V_N$ which accounts for the n-type conductivity of as-grown GaN. $V_{Ga}$ has a deep acceptor level. $Ga_i$ is a donor with its donor level in the conduction band. Split $N_i$ has a deep acceptor level. Neutral $N_{Ga}$ has an energy level near the middle and thus is stable in the middle of the band gap. Losing or gaining electrons introduces shallow acceptor or donor levels in the band gap. This character of $N_{Ga}$ makes it a plausible candidate for the experimentally observed broad red and green luminescence that detected only in high-resistivity GaN samples. We also give formation energies of each defect, among which vacancies are the most favourable ones in different conditions. We calculate ionization energies of different charge state of different defects and explain different photoluminescence.

**Acknowledgements**

ZX thanks the China Scholarship Council (CSC) for support. We are grateful to Alex Ganose and Stephen Shevlin for technical help, and Chris Van de Walle and Audrius Alkauskas for useful discussions. The EPSRC is acknowledged for funding (EP/K038419; EP/I03014X; EP/K016288). AW was supported by the Royal Society. Computational resources were provided through the Materials Chemistry Consortium on EPSRC grant number EP/L000202.

# Donor and Acceptor Characteristics of Native Point Defects in GaN

Zijuan Xie[1,2], Yu Sui[*,1], John Buckeridge[†,2], C. Richard A. Catlow[2], Thomas W. Keal[3], Paul Sherwood[3], Aron Walsh[4,5], Matthew R. Farrow[2], David O. Scanlon[2,6,7], Scott M. Woodley[2], and Alexey A. Sokol[‡,2]

[1] Department of Physics, Harbin Institute of Technology, Harbin, China
[2] Kathleen Lonsdale Materials Chemistry, Department of Chemistry, University College London, London, United Kingdom
[3] Daresbury Laboratory, Scientific Computing Department, STFC, Daresbury, Warrington, United Kingdom
[4] Department of Materials, Imperial College London, London, United Kingdom
[5] Department of Materials Science and Engineering, Yonsei University, Seoul, Korea
[6] Thomas Young Centre, University College London, London, United Kingdom
[7] Diamond Light Source Ltd., Diamond House, Harwell Science and Innovation Campus, Didcot, Oxfordshir, United Kingdom

E-mail: [*]suiyu@hit.edu.cn, [†]j.buckeridge@ucl.ac.uk, [‡]a.sokol@ucl.ac.uk

**Supplemental Material**

## 1. Calculation of defect formation energies

The method we have used to determine defect formation energies of native defects in the manuscript, is equivalent to calculating the energies of corresponding defect reactions. We distinguish here the condition of 'anion-rich' and 'anion-poor' used in the manuscript, follow the standard Kröger-Vink notation, i.e. superscript indices on the right-hand side of the defect symbol indicate its charge state: the cross ($^\times$) means neutral, the prime ($'$) – singly negatively charged, and the dot ($^\bullet$) – singly positively charged species. We present below the defect reactions for each native defect studied in the manuscript in both conditions, when all defects are in their neutral charge state.

For N vacancy:

$$N_N^\times \rightarrow V_N^\times + \tfrac{1}{2} N_2(g) \quad \text{(anion-rich)}$$
$$N_N^\times + Ga(s) \rightarrow V_N^\times + GaN(s) \quad \text{(anion-poor)}$$

For Ga vacancy:

$$Ga_{Ga}^\times + \tfrac{1}{2} N_2(g) \rightarrow V_{Ga}^\times + GaN(s) \quad \text{(anion-rich)}$$
$$Ga_{Ga}^\times \rightarrow V_{Ga}^\times + Ga(s) \quad \text{(anion-poor)}$$

For Ga interstitial:

$$Ga_{Ga}^\times + GaN(s) \rightarrow Ga_i^\times + \tfrac{1}{2} N_2(g) \quad \text{(anion-rich)}$$
$$Ga_{Ga}^\times + Ga(s) \rightarrow Ga_i^\times \quad \text{(anion-poor)}$$

For N interstitial:

$$\text{Ga}_{\text{Ga}}^{\times} + \tfrac{1}{2}\text{N}_2(g) \rightarrow \text{N}_i^{\times} \qquad \text{(anion-rich)}$$
$$\text{Ga}_{\text{Ga}}^{\times} + \text{GaN}(s) \rightarrow \text{N}_i^{\times} + \text{Ga}(s) \qquad \text{(anion-poor)}$$

For N antisite:

$$\text{Ga}_{\text{Ga}}^{\times} + \text{N}_2(g) \rightarrow \text{N}_{\text{Ga}}^{\times} + \text{GaN}(s) \qquad \text{(anion-rich)}$$
$$\text{Ga}_{\text{Ga}}^{\times} + \text{GaN}(s) \rightarrow \text{N}_{\text{Ga}}^{\times} + 2\text{Ga}(s) \qquad \text{(anion-poor)}$$

In these reactions, we consider molecular nitrogen, $N_2$ gas, solid GaN and / or solid metallic Ga as the source or sink for elemental nitrogen and gallium in their standard state (at room temperature and ambient pressure), the relevant enthalpies of which are further corrected to the absolute zero temperature limit. We calculate the heat of formation of an $N_2$ molecule as -109.529 Ha using the B97-2 exchange and correlation density functional and -109.521 Ha using the BB1k functional, while also employing the same basis as the one used in our QM/MM cluster calculations. The heat of formation of solid GaN is -110.5 kJ/mol[1]. The heat of atomisation of solid Ga is 272 kJ/mol[1]. In our basis set, as described in reference 32 of the paper, we have removed outer diffuse functions on Ga, which allowed us to reduce the size of the QM/MM model and therefore computational costs while retaining the accurate reproduction of Ga ions in the high charge states (+2 and +3). Consequently, the description of neutral atoms using these basis sets would be too poor. As the ionization potentials were part of the training set of thermochemical properties, along with the binding energies, for the hybrid functionals we use, the experimental and calculated values should be in agreement. Thus, we calculate the Ga atom in +3 charge state with the same basis sets and pseudopotential, and get the energy of -257.681 Ha using the B97-2 and -257.417 Ha with the BB1k functional. Then we take into account the 1$^{st}$, 2$^{nd}$, and 3$^{rd}$ ionisation potentials of Ga, which are 5.999 eV, 20.515 eV and 30.726 eV, respectively[1]. Finally, we obtain the energy of the neutral Ga atom of -259.785 Ha using the B97-2 and -259.521 Ha with the BB1k functionals. Thus, in addition to the defect energies, having the energy of a Ga atom, the heat of formation of $N_2$ and the heat of formation of GaN, we are able to determine all energies necessary for the reactions listed above.

## 2. Calculated bond lengths in GaN

In table S1 we compare the Ga-N bond lengths determined using (i) our interatomic potential model (MM in the table)[2,3]; (ii) density functional theory employing the B97-2 hybrid functional within our QM region; (iii) density functional theory employing the BB1k hybrid functional within our QM region; and (iv) XRD measurements at 100 K on GaN layers grown on sapphire[4] (as reported for an exposed layer possibly containing defects; the corresponding value from the buffer layer is given here and below in parentheses), and at 294 K on powder samples - see [5] and references therein.

Table S1. Ga-N bond lengths (axial and basal) from our MM model, at the centre of our embedded cluster employing both functionals, and from experiment.

| Ga-N bond lengths Å | MM[2,3] | B97-2 | BB1k | Experiment [4,5] |
|---|---|---|---|---|
| axial | 1.942 | 1.941 | 1.931 | 1.9664 (1.9582) |
| basal | 1.951 | 1.963 | 1.949 | 1.9527 |

## 3. Interatomic potential model

In table S2, we present structural ($a$, $c$ and $u$), elastic ($c_{ij}$), dielectric ($\varepsilon_0^{ij}, \varepsilon_\infty^{ij}$) and lattice dynamical ($v$) properties of GaN in the wurtzite and zinc blende phases, as calculated using our interatomic potential model and compared with experiment[6] and a previously derived potential model[7]. See Ref. [2,3] and references therein for further details. Note that in [2,3] two sets of interatomic potentials have been derived, one specifically for GaN intended for hybrid QM/MM calculations and another for solid solutions in the III-V nitride series of compounds.

Table S2. Calculated properties of wurtzite and zinc-blende GaN by our interatomic potential model in comparison with room-temperature experimental data [4–8] and a previous potential model [7]. Note that there is paucity of experimental data in the literature for the zinc blende (cubic) phase of GaN and therefore we provide comparison only for the wurtzite phase.

| Property | Wurtzite | | | Zinc blende | |
|---|---|---|---|---|---|
| | Experiment [4–8] | Present potential | Zapol *et al.* [7] potential | Present potential | Zapol *et al.* [7] potential |
| $a$ (Å) | 3.1972; 3.1894 | 3.1863 | 3.23 | 4.509[a] | |
| $c$ (Å) | 5.207 (5.186); 5.1861 | 5.1845 | 5.16 | | |
| $u$ | 0.3776; 0.3789 | 0.3747 | 0.385 | | |
| $c_{11}$ (GPa) | 360 | 375 | 386 | 379 | 300 |
| $c_{33}$ (GPa) | 392 | 496 | 391 | | |
| $c_{12}$ (GPa) | 130 | 238 | 160 | 216 | 191 |
| $c_{13}$ (GPa) | 105 | 227 | 141 | | |
| $c_{44}$ (GPa) | 100 | 72 | 115 | 74.1 | 160 |
| $c_{66}$ (GPa) | 115 | 69 | 113 | | |
| $\varepsilon_0^{11}$ | 9.38 | 10.01 | 8.05 | 9.7 | 8.88 |
| $\varepsilon_0^{33}$ | 10.2 | 10.25 | 11.2 | | |
| $\varepsilon_\infty^{11}$ | 5.35 | 5.34 | 5.21 | 5.3[b] | 5.41 |
| $\varepsilon_\infty^{33}$ | 5.35 | 5.43 | 5.84 | | |
| $v_{low}$ (E$_2$) (cm$^{-1}$) | 144 | 109 | | 555 | |
| $v_{high}$ (E$_1$) (cm$^{-1}$) | 744 | 746 | | 751 | |

[a] Experimental values from strained samples grown on various substrates are in the range of 4.49 – 4.55 Å at room temperature[9,10].
[b] Experimental value of 5.86 is obtained from saturated refraction index measurements[11].

4. **Summary of defect states**

We discuss here the electronic states that result on defect formation, as well as structural effects where relevant. We refer to a state as stable if it possesses a donor (acceptor) level not resonant in the conduction (valence) band and hence is not subject to autoionization. However, a donor (acceptor) level that lies in the valence (conduction) band, while referred to here as stable, strictly speaking represents a local deformation of the valence (conduction) band around the defect and hence can only be observed in macroscopic samples as peaks associated with the band edge. Such levels are indicated by thicker lines in the vertical transition energies diagrams below.

*A nitrogen vacancy, $V_N$,* is found to be stable in five charge states: (-1, 0, +1, +2, and +3). A missing $N^{3-}$ ion leaves the +3 state of $V_N$ with no electron remaining in the vacant site; the positive Madelung potential due to the vacancy therefore pushes electronic levels of the surrounding nitride ions below the VBM by *ca*. 1.71-1.72 eV as determined using both employed functionals – see figure S1. This triply positively charged vacancy can trap electrons and form bound states in the band gap. This behaviour is typical for isoelectronic F centres in halide and oxide systems, where the Madelung potential stabilises electrons at vacant anionic sites[12,13]. Trapping one or two electrons by the vacancy in the +3 charge state is expected to result in occupation of a hydrogenic 1s-like state centred on the vacant site. In the open-shell configuration of the +2 state, we do find the spin density of the electron mainly contained within the vacant site with outer, low-value isodensity surfaces noticeably deformed, adopting the symmetry of the hexagonal host lattice – as can be seen from figure S1.

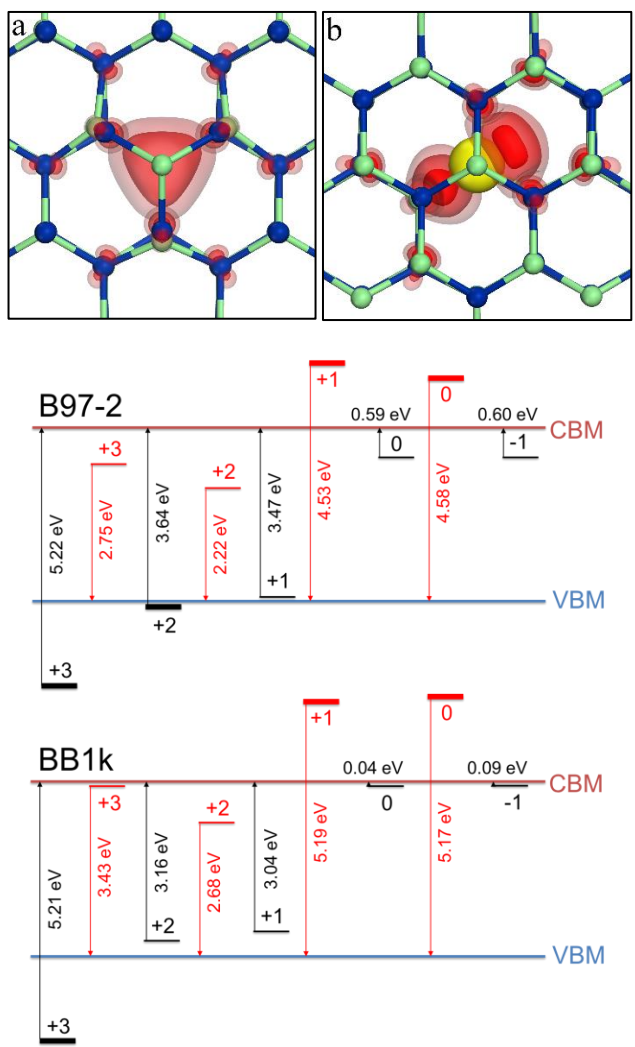

**Figure S1.** A nitrogen vacancy in GaN in charged states +2 (top left) and 0 (top right). The spin densities of the +2 and 0 states are indicated by isosurfaces 0.01, 0.005, 0.0025 au highlighted in red. Optical energy levels of $V_N$ with respect to band edges (vertical ionisation potentials with respect to vacuum), calculated using the B97-2 and BB1k density functionals are shown in the bottom part of the figure. Black lines denote electron ionization to the conduction band minimum (CBM), and red lines hole ionization to the valence band maximum (VBM). Here and in following figures, the charge state of the defect is indicated

above or below the corresponding level bar, and the thicker lines represent states that are either unstable (indicated by downwards-pointing black arrows or upwards-pointing red arrows) or form local distortions of the appropriate band edge in the vicinity of the defect.

The donor levels of the +2 and +1 states are close to each other in the band gap near the VBM, with a difference of about 0.1 eV. The closed shell $V_N^+$ state, with two trapped electrons, represents the limit to the number of electrons the vacancy can localise fully. In the neutral state, as two electrons occupy the vacancy 1s like state, the third extra electron is stabilised in two long Ga-Ga bonds with an antibonding contribution from the 2s state on the vacancy, and a σ-type Ga-Ga bonding state is based predominantly on ionic Ga 4sp shells and is curved around the central charge density spheroid formed by the first two electrons (see figure S1). The Ga-Ga bond is stabilised by the sizable mismatch of Ga and N ions making the distance between Ga ions here comparable to that in the Ga metal[14]. The energy levels of the neutral and the -1 state thus are close to each other and the conduction band minimum (CBM), which means that $V_N$ is a shallow donor in GaN, as identified in experiments[15,16]. The result for the ionisation from the -1 to 0 charge state differs from our previous report in Ref. [17], becoming 0.04 eV deeper, due to an improved account of the high-frequency dielectric response (see section 2 in the paper), but this difference does not alter our conclusions regarding the shallow donor nature of N vacancies. On charge reduction, the Ga ions surrounding the vacancy move towards the vacancy site (as we add more electrons) because of the screening of the Coulomb repulsion, as summarised in table S3.

**Table S3.** Geometry of the nitrogen vacancy in GaN: nearest neighbour Ga ion displacements from the lattice sites relative to the corresponding Ga–N bond lengths (%), obtained using the B97-2 and BB1k density functionals. Negative (positive) numbers indicate contraction (expansion) of interionic distances for each defect charge state ($q$).

| $q$ | Axial neighbour | | | Axial+basal | Basal neighbour | | |
|---|---|---|---|---|---|---|---|
| | B97–2 | BB1k | sX-LDA [18] | HSE[19] | B97–2 | BB1k | sX-LDA [18] |
| +3 | 23 | 24 | 18.1 | 21.9 | 22 | 23 | 16.6 |
| +2 | 9 | 5 | | 10.8 | 12 | 13 | |
| +1 | -6 | -9 | 1.6 | 1.1 | 3 | 4 | 0.5 |
| 0 | -10 | -10 | | | -1 | -1 | |
| -1 | -9 | -10 | | | -4 | -4 | |

A gallium vacancy, $V_{Ga}$, is found to be stable in five charge states: (-3, -2, -1, 0, and +1). A missing $Ga^{3+}$ ion leaves the -3 state of $V_{Ga}$ with no electrons localised in the vacant site, as the Madelung potential destabilises electron states centred on vacant cationic sites. In the ground state, the three electrons donated by the missing Ga atom are found to occupy lone pair states centred on the three nearest neighbour basal N ions. In the absence of screening by the F-centre like charge density in the vacancy, these surrounding N ions repel each other strongly (by about a quarter of a Ga-N bond length in the perfect material) moving outwards of the vacancy and each other, as can be seen from table S4. The smaller displacements reported in Ref. [18] may be attributed to the elastic strain on the defect in supercell calculations due to the periodic boundary conditions.

The electron rich $V_{Ga}^{3-}$ state can lose electrons from the surrounding N lone pairs: starting successively with the basal and finishing with the axial N ions. Thus, in the -2 state, one of three basal N ions binds a hole, resulting in a spin doublet; in the -1 state, two of the basal N ions bind a hole each, forming a spin triplet; in the neutral state, all three basal N ions bind a hole, which is stable as a spin quadruplet. Those N ions that bind holes relax more than those do not. This trend has also been observed from hybrid (HSE) supercell calculations[20]. Clark *et al.*[21] reported about 0.5 eV gain in energy due to localisation by hybrid functionals compared to the LDA for wurtzite structured wide-gap semiconductor ZnO. This stabilisation has also been seen in other materials, like $TiO_2$[22] and $MoS_2$[23]. All the above charge states introduce donor energy levels in the bandgap, whereas losing another electron from the axial N lone pair pushes unpaired, locally highest occupied electron states into the valence band by *ca.* 0.4 eV, as shown in figure S2. In the neutral state, the BB1k functional gives a stronger hole localization than B97-2. From the BB1k calculation, we find that the $V_{Ga}^0$ state binds a hole on each basal N ion with a spin population of ca. 1.0, and on the axial N ion with a population below 0.1. Using the B97-2 functional, the spin populations are 0.8 on the basal N and 0.7 on the axial N ions.

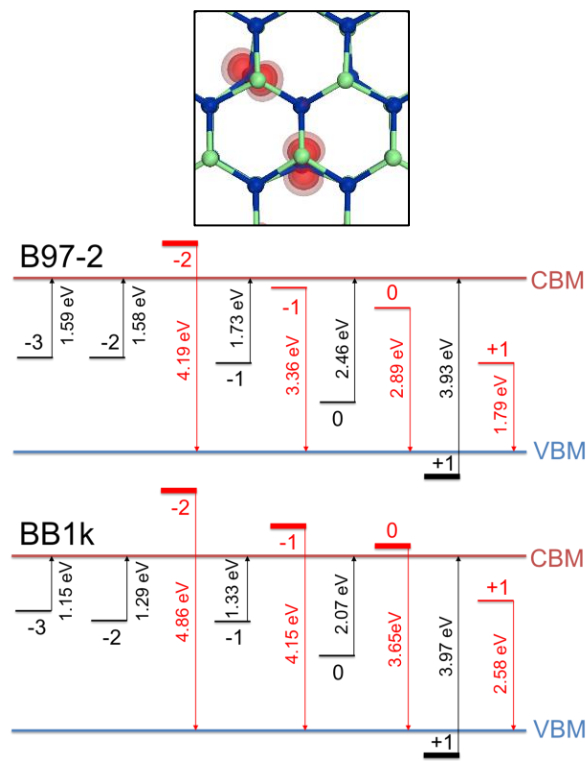

**Figure S2.** (a) Ga vacancy in GaN in charge state -1. Spin density distribution of the triplet state is indicated by (red) isosurfaces 0.05, 0.025, 0.01 au. (b) Optical energy levels (in eV) of $V_{Ga}$ with respect to band edges (vertical ionisation potentials with respect to vacuum), calculated using the B97-2 and BB1k density functionals.

**Table S4**. Geometry of the gallium vacancy in GaN: nearest neighbour N ion displacements from the lattice sites relative to the corresponding Ga–N bond lengths (%), obtained using the B97-2 and BB1k density functionals for each defect charge state ($q$).

| $q$ | Axial neighbour | | Basal neighbours without holes | | Basal neighbours binding holes | |
|---|---|---|---|---|---|---|
| | B97–2 | BB1k | B97–2 | BB1k | B97–2 | BB1k |
| -3 | 5 | 6 | 12 | 13 | - | - |
| -2 | 5 | 4 | 10 | 11 | 15 | 16 |
| -1 | 4 | 3 | 8 | 9 | 14 | 15 |
| 0 | 5 | 1 | - | - | 11 | 15 |
| +1 | 9 | 8 | - | - | 14 | 14 |

*A gallium interstitial*, $Ga_i$ stabilizes at an octahedral interstitial site in the hexagonal channel of GaN, in a "cage" composed by three Ga ions in the basal (0001) plane and three N ions in another basal (0001) plane both up and down the *c* axis, showed in figure S3. $Ga_i$ stabilises close to the nearest N plane due to the Coulomb attraction, and the distances between $Ga_i$ and the three N ions are comparable to a typical Ga–N bond length in the lattice. The $Ga_i^{3+}$ state is similar to a Ga ion in the lattice; thus it stabilizes in the centre of the channel, pushing the six nearest Ga ions away with 9%-16% displacements compared to a typical Ga–N bond length in the lattice, while the surrounding N ions relax to accommodate it. As with other electron-poor positively charged defects, the local highest occupied states are pushed below the top of the valence band, for this species by about 1.64-1.69 eV. This +3 state will act as a resonance in the valence band of the crystal, and can attract two electrons to a compact Ga-4*sp*-like lone pair state in the band gap (see figure S3). However, the electrons trapped on the $Ga_i$ ion are somewhat delocalized. Furthermore, in the neutral state, $Ga_i$ introduces a compact resonance in the conduction band and therefore behaves as a shallow donor stabilising effective-mass states. Both in the neutral and +1 state, we observe a large distortion: the interstitial Ga atom moves off the channel centre by 26% (+1 state) and 34% (neutral state) of a typical Ga–N bond length (1.948 Å), see figure S3c. Such a distortion has been reported recently by Kyrtsos *et al* based on HSE calculations[24]. In the neutral state, the third electron trapped on the $Ga_i$ ion is also not fully localized, with only ca. 0.5 spin population on the interstitial ion.

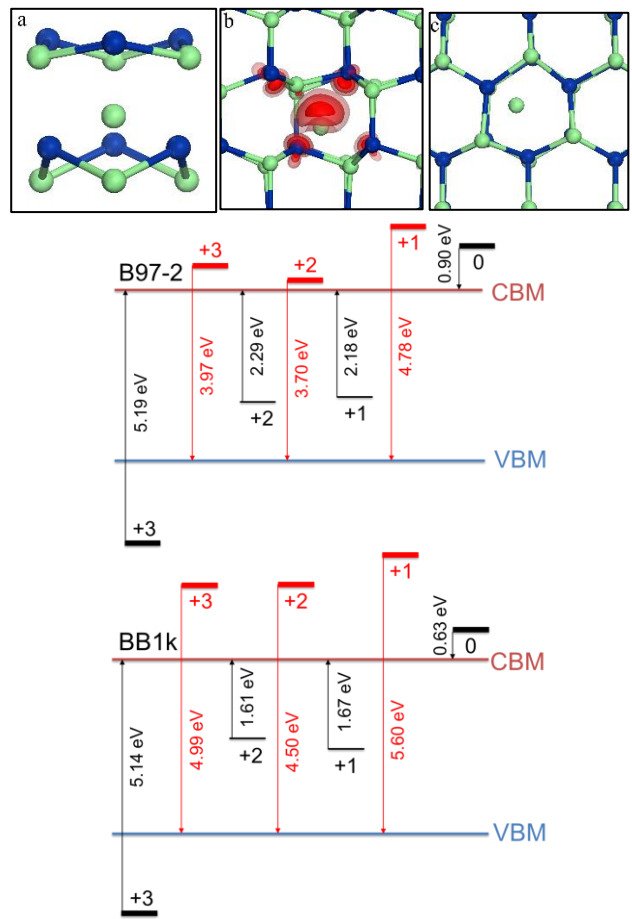

**Figure S3.** Ga interstitial in GaN in charge states +3 (a), +2 (b) and +1 (c). Spin density of +2 state is indicated by (red) isosurfaces of 0.02, 0.01, 0.005 au. Optical energy levels (in eV) of $Ga_i$ with respect to band edges (vertical ionisation potentials with respect to vacuum),

calculated using the B97-2 and BB1k density functionals are shown in the bottom part of the figure.

*A nitrogen interstitial, $N_i$*, stabilizes in two alternative configurations: at an octahedral interstitial site in the hexagonal channel as $Ga_i$ described above, and as a split interstitial forming a dumbbell structure reminiscent of the well-known $V_k$ centres in alkali metal halides, peroxy and superoxide dioxygen species in oxides, or simply an $N_2$ molecule in the gas phase. $N_i$ as an octahedral interstitial is found to be stable in three charge states: (-3, -2, and -1). Similar to $Ga_i$, $N_i^{3-}$ resembles a $N^{3-}$ ion in the lattice and is located at the channel centre but closer to the nearest Ga-(0001) plane. Removing electrons from $N_i^{3-}$ results in high-spin quasiatomic states with defect levels in the bandgap – see figure S4. The first hole localises on a N 2p orbital aligned in the basal plane as can be seen from the classical dumbbell configuration (figure S4b); the second hole occupies another basal p orbital, and the spin density adopting a torus shape (figure S4c); finally, in the charge neutral state, the octahedral configuration becomes energetically unfavourable giving way to a split interstitial. The surrounding ions relax more with the charge state changing from -1 to -3, as shown in table S5. When negatively charged, the interstitial ion moves towards the Ga plane by 7% (-2 state) and 13% (-3 state) compared to Ga-N bond length (1.948 Å), with Ga ions moving slightly off-site and the closest N plane repelled strongly. The opposite N plane is affected only mildly.

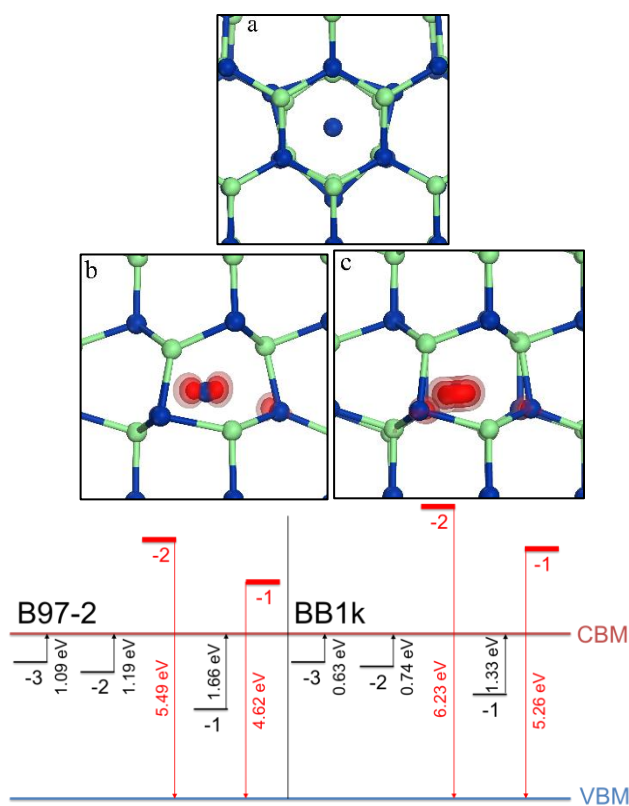

**Figure S4.** Octahedral N interstitial in GaN in charge states -3 (a), -2 (b) and -1(c). Spin density of the -2 state is indicated by (red) isosurfaces of 0.1, 0.05, 0.025 au. Spin density of the -1 state is indicated by (red) isosurfaces of 0.2, 0.1, 0.05 au. Optical energy levels (in eV) of $N_i$ with respect to band edges (using vertical ionisation potentials with respect to vacuum), calculated using the B97-2 and BB1k density functionals are shown in the bottom part of the figure.

**Table S5.** Geometry of octahedral nitrogen interstitial in GaN: displacements of nearest and next nearest neighbour ions from the lattice sites as a percentage of average Ga–N bond lengths in GaN (1.948 Å), obtained using the B97-2 and BB1k density functionals for each charge state $q$. Upper and lower plane N are assigned with reference to the crystallographic $c$ axis. Note that using either functional results in similar relaxations. The minus sign indicates a bond contraction.

| $q$ | Upper plane N | Lower plane N | Ga |
|---|---|---|---|
| -3 | 19%~20% | 5% | -(7%~12%) |
| -2 | 16%~21% | 1%~5% | -(1%~9%) |
| -1 | 14%~15% | 1%~2% | -(1%~4%) |

$N_i$ as a split interstitial is found to be stable in five charge states: -2, -1, 0, 1, and 2; and two characteristic geometries depending on the orientation of the N dimer with very small energy differences between the two in negative and neutral states. We show the two geometries in figure S5a and b, where a molecular ion aligns in the $<1\overline{3}0>$ and $<0\overline{3}1>$ directions, respectively. The total energies in the two configurations are very similar, differing by less than 0.16 eV per cluster. The electronic configuration of $N_2$ corresponds to the defect in the charge state 3+. The triple bond between the two N atoms makes them share the former lattice N ion position and act like a single lattice N ion. As we consider the evolution of the N-N bond on addition of electrons to the defect, we observe a gradual deconstruction of the bond from the strongest triple bond to a completely nonbonding configuration when an interstitial atom adopts an octahedral position. The added electrons occupy antibonding orbitals of the $N_2$ species with every two electrons per N atom making a (nonbonding) lone pair. In the neutral defect state for example, we find an effective 1.5 bonds remaining in the nitrogen dimer. In the -1 state, only a single bond will link two nitrogen atoms; and, in the -2 state, the bond is ruptured, leaving one electron, which is delocalized across the whole defect region as indicated by its spin density. We find the energy levels of $N_i^{2-}$ above the CBM, which indicates its resonance nature, i.e. that the defect would autoionize, spontaneously losing an electron to the conduction band. This analysis is supported by the trend of the split interstitial N-N bond length increasing with the number of electrons on the nitrogen dimer, as shown in table S6, and the spin density distribution in the open shell electronic configurations highlighted in figure S5. The neutral state spin density has been identified in experiments[25,26], and similar spin density distributions in the neutral and +1 charge states have been found in HSE calculation of wurtzite aluminium nitride[27].

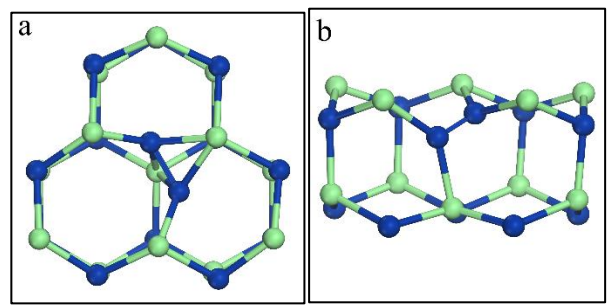

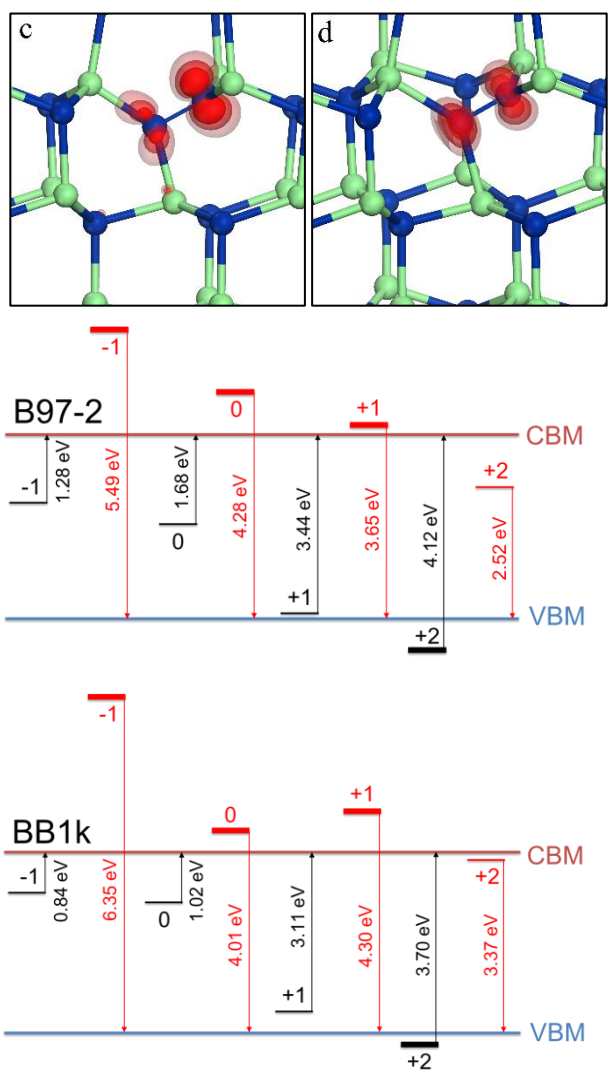

**Figure S5.** Split N interstitial in GaN in charge states -1 (a) and +1 (b). Spin density of the 0 (c) and +1 (d) states is indicated by (red) isosurfaces of 0.1, 0.05, 0.025 au. Optical energy levels (in eV) of $N_i$ with respect to band edges (using vertical ionisation potentials with respect to vacuum), calculated using the B97-2 and BB1k density functionals are shown in the bottom part of the figure.

**Table S6.** Nitrogen dimer bond lengths in the split nitrogen interstitial in GaN obtained using the B97-2 (BB1k) density functionals for each charge state $q$ – cf. the $N_2$ bond length of 1.098 Å.

| $q$ | N-N bond length (Å) | Other calculations (Å) |
|---|---|---|
| -2 | 1.56 (1.57) | |
| -1 | 1.56 (1.57) | 1.45[18], 1.41[28] |
| 0 | 1.34 (1.34) | 1.40[29] ~1.33[18] |
| +1 | 1.26 (1.26) | ~1.24[18] |
| +2 | 1.17 (1.16) | ~1.18[18] |

*A nitrogen antisite*, $N_{Ga}$, in its neutral state forms a triple bond with its axial N neighbour (designated as $N_a$ here) like a gas-phase $N_2$ molecule, which resembles the configuration of the nitrogen split interstitial (see figure S6a). This triple bond is still along the former Ga-N bond but is much shorter with the $N_{Ga}$ atom moved down by 0.55 Å and the $N_a$ atom up by 0.26 Å, resulting in an N-N bond length close to that in an $N_2$ molecule (see table S7). This defect can therefore be considered as a defect complex of an $N_2$ interstitial straddling adjacent Ga and N vacancies in their formal charge states. When adding electrons to the neutral state, the two N ions move concertedly along the c axis toward the three positive basal Ga ions around the regular N lattice site, approaching the vacancy (see figure S6c), while the N-N bond distance expands by ~10%. Similar geometry was seen in references [30,31]. The tendency of the N ion to move off the Ga lattice site can be attributed to a size mismatch. The electron added to the complex localises on the $N_2$ interstitial breaking one of the covalent bonds, hence the bond elongation. On adding a second electron, an $O_2$ like triplet electronic configuration is stabilised. An alternative configuration of the antisite is triply degenerate with the $N_2$ dimer orienting along one of the Ga-N basal bonds, see figure S6e. The calculated defect energies of these two configurations are again very similar (cf. nitrogen split interstitials).

Upon removal of one electron from the neutral defect, a hole is localised on the $V_{Ga}$, for which the BB1k XC density functional predicts a more localised character (spin population of 1.0) than the B97-2 (spin population of 0.5), and the $N_2$ interstitial moves along the axis towards the three negative basal N ions around the vacancy. The geometry of the +1 and neutral states are similar, but the Ga-N bond involving the N ion that binds a hole expands by 2%, see figure S6b. Removing a second electron, rather than localising a second hole in the $V_{Ga}$, the $N_2$ interstitial moves towards the $V_{Ga}$ and the electronic configuration reconstructs with one of the covalent bonds broken, resulting in an elongated N-N bond length, a stable trigonal structure (see figure S6g) and no holes in the vacancy. On further ionisation, the energy levels lower into the valence band as the third hole localises again on the same basal N ion in $V_{Ga}$, as can be seen in figure S6h, while the trigonal configuration is retained across the vacancy.

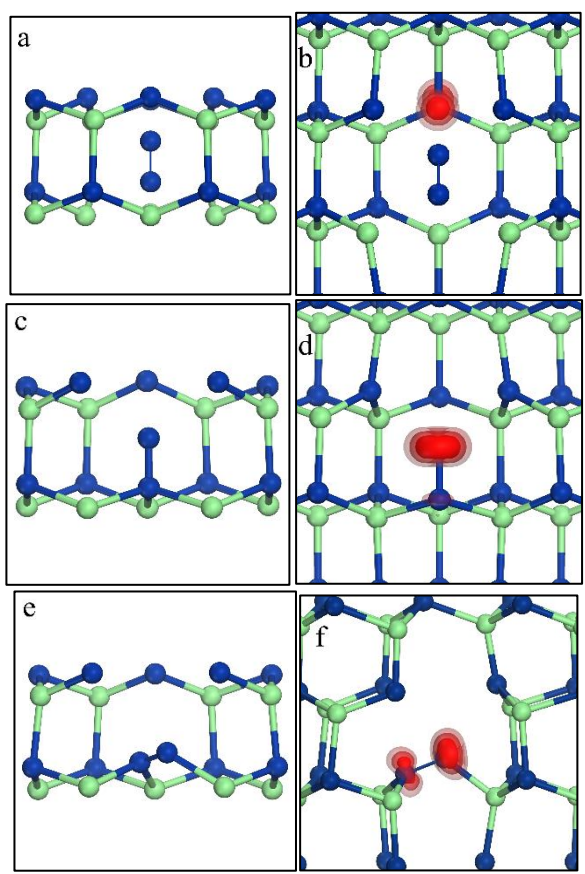

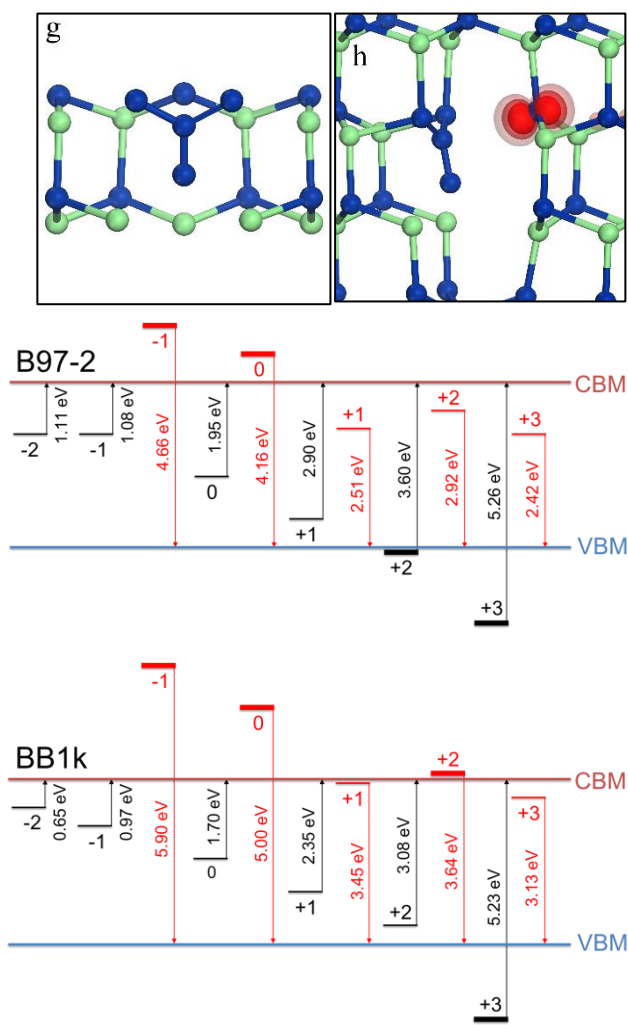

**Figure S6.** N antisite in GaN in charge states 0 (a), +1 (b), -2 (c-f), +2 (g) and +3 (h). Spin density is indicated by (red) isosurfaces of 0.1, 0.05, 0.025 au. Optical energy levels (in eV) of $N_{Ga}$ with respect to band edges (using vertical ionisation potentials with respect to vacuum), calculated using the B97-2 and BB1k density functionals are shown in the bottom part of the figure.

**Table S7.** Nitrogen dimer bond lengths at the nitrogen $N_{Ga}$ antisite in GaN obtained using the B97-2 and BB1k density functionals for each charge state *q*.

|     | N-N bond length (Å) | | |
| --- | --- | --- | --- |
| *q* | B97-2 | BB1k | Reference |
| -2 | 1.26 | 1.26 | |
| -1 | 1.23 | 1.19 | |
| 0 | 1.16 | 1.12 | 1.31[30], 1.10[29] |
| +1 | 1.18 | 1.11 | |
| +2 | 1.28 | 1.26 | |
| +3 | 1.29 | 1.27 | |

In this work, we have not considered a $Ga_N$ antisite, as discussed in the paper section 3.1.